\documentclass[aps,prl,twocolumn,superscriptaddress]{revtex4-1}

\usepackage{arydshln}
\usepackage{graphicx}
\usepackage{dcolumn}
\usepackage{enumerate,amsthm,amssymb,color}
\usepackage{bbm}
\usepackage{bm}
\usepackage{multirow}
\usepackage{url}
\usepackage{breakurl}
\usepackage[breaklinks]{hyperref}
\usepackage[french, english]{babel}
\usepackage{synttree}
\usepackage{float}
\usepackage{newfloat}
\usepackage{mathtools}
\DeclareFloatingEnvironment[
    fileext=los,
    listname=List of Schemes,
    name=Supplementary Figure,
    placement=H,
    within=section,
]{supfigure}
\DeclareFloatingEnvironment[
    fileext=los,
    listname=List of Schemes,
    name=Supplementary Protocol,
    placement=H,
    within=section,
]{supprotocol}
\DeclareFloatingEnvironment[
    fileext=los,
    listname=List of Schemes,
    name=Supplementary Table,
    placement=H,
    within=section,
]{suptable}
\usepackage{adjustbox}

\usepackage{booktabs}
\usepackage{mdframed}
\usepackage{arydshln}
\usepackage{graphicx}
\usepackage{dcolumn}
\usepackage{amsmath}
\usepackage{synttree}
\usepackage{float}
\usepackage{newfloat}
\DeclareFloatingEnvironment[
    fileext=los,
    listname=List of Schemes,
    name=Box,
    placement=H,
    within=section,
]{protocol}
\usepackage{adjustbox}

\newcommand{\norm}[1]{\left|\left|#1\right|\right|}

\newcommand{\ket}[1]{|#1\rangle}
\newcommand{\bra}[1]{\langle#1|}
\newcommand{\ketbra}[2]{|#1\rangle\langle#2|}

\newcommand{\altketbra}[1]{|#1\rangle \langle #1|}

\newcommand{\COMMENT}[1]{}

\graphicspath{{images/}}

\begin{document}


\title{Experimental plug and play quantum coin flipping}

\author{Anna Pappa}\affiliation{LTCI, CNRS - T\'el\'ecom ParisTech, Paris, France}\affiliation{LIAFA, CNRS - Universit\'e Paris Diderot 7, Paris, France}
\author{Paul Jouguet}\affiliation{LTCI, CNRS - T\'el\'ecom ParisTech, Paris, France}\affiliation{SeQureNet, Paris, France}
\author{Thomas Lawson}\affiliation{LTCI, CNRS - T\'el\'ecom ParisTech, Paris, France}
\author{Andr\'e Chailloux}\affiliation{INRIA Paris-Rocquencourt, France}
\author{Matthieu Legr\'e}\affiliation{ID Quantique SA, Geneva, Switzerland}
\author{Patrick Trinkler}\affiliation{ID Quantique SA, Geneva, Switzerland}
\author{Iordanis Kerenidis}\affiliation{LIAFA, CNRS - Universit\'e Paris Diderot 7, Paris, France}\affiliation{Center for Quantum Technologies, National University of Singapore, Singapore}
\author{Eleni Diamanti}\affiliation{LTCI, CNRS - T\'el\'ecom ParisTech, Paris, France}

\begin{abstract}
\noindent{Performing complex cryptographic tasks will be an essential element in future quantum communication networks. These tasks are based on a handful of fundamental primitives, such as coin flipping, where two distrustful parties wish to agree on a randomly generated bit. Although it is known that quantum versions of these primitives can offer information-theoretic security advantages with respect to classical protocols, a demonstration of such an advantage in a practical communication scenario has remained elusive. Here we experimentally implement a quantum coin flipping protocol that performs strictly better than classically possible over a distance suitable for communication over metropolitan area optical networks. The implementation is based on a practical plug and play system, developed by significantly enhancing a commercial quantum key distribution device. Moreover, we provide combined quantum coin flipping protocols that are almost perfectly secure against bounded adversaries. Our results offer a useful toolbox for future secure quantum communications.
}\clearpage
\end{abstract}

\date{\today}

\maketitle
\clearpage

\section{Introduction}
Security is an imperative in all communication networks. Quantum communications hold the promise of achieving a security level that is impossible to reach by purely classical means. Indeed, information-theoretic security has been demonstrated for the cryptographic task of distributing a secret key between two trusted and collaborating communicating parties using systems exploiting quantum effects \cite{SBC:rmp09}. However, many advanced cryptographic schemes belong to a model where the two parties do not trust each other and hence cannot collaborate. One of the fundamental primitives in this setting is coin flipping, in which two spatially separated distrustful parties share a randomly generated bit, whose value must be unbiased \cite{Blum:ece81}.
This functionality, either between two or more parties, is used in communication networks, for instance in online gaming and in randomized consensus protocols (due to its equivalence to the leader election functionality, a fundamental primitive in distributed computing), and it is also an integral component for secure function evaluation \cite{goldreich01}. When multiple parties are involved, coin flipping can be securely performed when more than half of the parties are honest. However, for two parties, which is the case that interests us here, we do not have such an honest majority. It is known that in the asynchronous classical model, perfectly secure coin flipping, i.e., with a zero bias, is impossible without computational assumptions, while in the synchronous (or relativistic) model, unconditionally secure perfect coin flipping is possible \cite{Kent:prl99}, at the expense of complex spatial configuration restrictions. Unfortunately, the impossibility result in the non-relativistic setting that we consider in this work holds even when protocols are enhanced with quantum communication \cite{LC:physicad98,May:prl97}. A series of theoretical works, however, have demonstrated that the probability that an all-powerful malicious party can bias the coin, namely the cheating probability, can be strictly lower than 1 in the quantum setting, with an ultimate asymptotic bound of $1/\sqrt{2}$ \cite{ATVY:stoc00,SR:prl02,Kit:qip03,Abdr:ccc04,NS:pra03,Amb:jcss04,CK:focs09}. Moreover, a weaker version of coin flipping, which still remains very useful for communication systems, can be almost perfectly achieved with quantum communication \cite{Moc:focs04,acgkm:arxiv14}. Coin flipping therefore provides a suitable framework to demonstrate an advantage of quantum over classical communication, achieving information-theoretic security in a non-cooperative model that is crucial for cryptographic applications beyond key distribution.

To demonstrate such an advantage, we need to consider all imperfections that naturally appear in practical devices. For photonic systems, which constitute the chosen architecture for quantum communications, imperfections typically appear in the form of losses in the channel and measurement apparatus, and errors in the different implementation stages. Furthermore, systems suitable for long-distance communications over fiber-optic channels usually employ coherent light sources, thus becoming vulnerable to attacks exploiting the non-deterministic photon emission inherent in such sources \cite{BLMC:prl00}.

Some of the aforementioned practical issues have been addressed in recent theoretical and experimental studies. An elegant solution to the problem of loss tolerance, i.e., the tolerance to photon losses at any communication distance, was presented in \cite{BBBG:pra09}, which however did not account for the presence of multi-photon pulses in coherent light source implementations. The cheating probability achieved by this protocol was slightly improved in subsequent work \cite{Cha:aqis10,AMS:pra10}. As a way to account for errors, the related primitive of bit string generation was also considered \cite{Kent:qcmc02,BM:pra04,LBA:PRL04}. In practice, a first implementation concerned a protocol that becomes insecure for any loss \cite{MVUZ:prl05}, while a promising solution gave results that unfortunately cannot be used in realistic conditions \cite{NFPM:njp08}. More recently, an implementation of the loss-tolerant protocol \cite{BBB:natcomm11} used an entangled-photon source to eliminate the problem of multi-photon pulses. This was the first experiment that demonstrated an advantage of quantum over classical communication for coin flipping in the presence of losses and errors. However, although in principle the cheating probability bound in the implemented protocol is independent of losses, a gain was shown in practice for a distance of a few meters. The closely related primitives of quantum bit commitment and oblivious transfer were experimentally demonstrated in the noisy storage model, where adversaries have access to an imperfect quantum memory \cite{NJM:natcom12,ENG:arxiv13}; however, these protocols do not offer security against all-powerful adversaries. Finally, quantum bit commitment with relativistic constraints was also recently implemented \cite{Lunghi:prl13,Liu:prl14}.

\begin{protocol*}[t!]
\begin{adjustbox}{minipage=1\linewidth,fbox,center}
\begin{tabular}{p{6.7cm}p{3.3cm}p{6.7cm}}
\hspace{1in}\framebox{\bf Alice}  &  & \hspace{1in}\framebox{\bf Bob}\\

\multirow{4}{6.5cm}{For each $i=1,\dots,K$, Alice randomly picks a basis $\alpha_i \in \{0,1\}$ and a
bit $c_i \in \{0,1\}$. \\
She sends to Bob $K$ pulses in states $|\Phi_{\alpha_i,c_i}\rangle$, where $\ket{\Phi_{\alpha_i,0}}=\sqrt{y}\lvert0\rangle+(-1)^{\alpha_i}\sqrt{1-y}\lvert1\rangle$, $\ket{\Phi_{\alpha_i,1}}=\sqrt{1-y}\ket{0}-(-1)^{\alpha_i}\sqrt{y}\ket{1}.$}   &   &  \\
  & ~~~~ $ \xrightarrow{\makebox[2.3cm]{$\ket{\Phi_{\alpha_i,c_i}}$}}$   &

 \multirow{2}{6.5cm}{For each $i=1,\dots,K$, Bob randomly picks a basis $\beta_i \in \{0,1\}$ and measures the  $K$ pulses in the bases $\{ \ket{\Phi_{\beta_i,0}}, \ket{\Phi_{\beta_i,1}} \}$.\\
Let $j$ be the position of the first measured pulse and $b_j$ the outcome of the measurement.\\
Bob sends to Alice $j$ and a random bit $b$.}\\

  &  & \\
 & \vspace{-0.15cm} $~~~~\xleftarrow{\makebox[2.3cm]{$j,~b\in\{0,1\}$}}$ &  \\
  \vspace{-0.3in}
Alice sends the basis $\alpha_j$ and the bit $c_j$ used for the $j$-th pulse.& \\
&  \vspace{-0.4in} ~~~~$\xrightarrow{\makebox[2.3cm]{$\alpha_j,~c_j$}}$&
\vspace{-0.4in}
 If the bases agree, $\alpha_j=\beta_j$, and the outcomes do not, $b_j \neq c_j$, then Bob aborts. Otherwise, the outcome of the coin value is $x=c_j\oplus b$.
\end{tabular}
\end{adjustbox}
\caption{{\bf Basic quantum coin flipping protocol.}}\label{protocol}
\end{protocol*}

Here, we provide a complete theoretical and experimental framework for the implementation of quantum coin flipping in practical communication scenarios. The protocol that we consider \cite{PCD:pra11} takes standard experimental imperfections (multi-photon emission, transmission loss, detector inefficiency and dark counts) into account. We show that our protocol can be combined with protocols that achieve almost perfect security, i.e., a bias asymptotically close to zero, against adversaries with bounded resources. More explicitly, if the adversary is bounded, then the protocol guarantees almost perfect security, while in the case of an all-powerful adversary, the protocol still guarantees a security level strictly higher than classically possible. Providing security against adversaries of varying complexity is of importance in the context of current communication networks, where technological and computational capabilities can evolve very rapidly. Furthermore, we experimentally implement the protocol using a practical plug and play system, developed by significantly enhancing a commercial quantum key distribution (QKD) device \cite{IDQ,SGG:njp02}. The key element of our implementation is that we take a realistic approach: to account for the unavoidable errors in the system and for coherent light source emission statistics, we allow for a non-zero but small probability of abort when both parties are honest, and accept the dependence of the cheating probability on communication loss thus departing from absolute loss tolerance \cite{BBBG:pra09,BBB:natcomm11}. This constitutes an important change with respect to previous protocols and leads to a gain of three orders of magnitude in communication distance. Indeed, using a security analysis pertaining to our implementation and an appropriate benchmark for classical coin flipping protocols \cite{HW:tcc11}, we can rigorously quantify the advantage offered by quantum communication as a function of distance, much in the way that the secret key fraction is calculated in practical QKD implementations \cite{SBC:rmp09}. In this way, we demonstrate a clear advantage for quantum coin flipping with information-theoretic security, at a communication distance suitable for metropolitan area network communications, with a system that can readily be deployed in such networks.


\section {Results}
\subsection{Basic quantum coin flipping protocol}

The protocol that we analyze and implement in this work is schematically shown in Box \ref{protocol} \cite{PCD:pra11}. Alice sends to Bob a fixed number $K$ of photon pulses in states $\ket{\Phi_{\alpha_i,c_i}}$, each of which is prepared independently following a uniformly random choice of basis $\alpha_i$ and bit $c_i$, with $i=1,\dots,K$, and a fixed protocol parameter $y$ (see Supplementary Figure 1 for a description of the states).  Bob measures the $K$ pulses by selecting uniformly at random bases $\beta_i$, and replies with the position of the first successfully measured pulse $j$ and a random bit $b$. Alice then reveals the basis and the bit used for that position: if the bases of the two parties agree, but the measurement output of Bob is not the same as Alice's bit, they abort. In all other cases, they agree that the coin value is $c_j \oplus b$.

A crucial feature of the protocol is the assumption that the states are generated by an attenuated coherent light source, which is a standard element of practical implementations. Therefore, each pulse contains a number of photons that follows a Poisson distribution with mean photon number $\mu$. Standard experimental imperfections are accounted for by introducing a probability to abort even when both parties are honest. This probability, denoted $H$, is determined by the experimental parameters, namely the mean photon number per pulse $\mu$, the number of protocol rounds $K$, the channel length, the detector quantum efficiency and dark count rate, and the error rate. Then, optimal cheating strategies for an all-powerful malicious party can be devised for both Alice and Bob, leading to expressions for the maximal cheating probabilities, $p^A_{q}$ and $p^B_{q}$, respectively (see Supplementary Note 1 for details). These are functions of $\mu$ and $K$, hence for a given desired honest abort probability, it is possible to minimize the cheating probabilities by finding optimal values for these parameters. Additionally, the parameter $y$ can be appropriately adjusted so that $p^A_q = p^B_q \equiv p_q$, which means that the protocol is fair.

\begin{figure*}
\includegraphics[width=17cm]{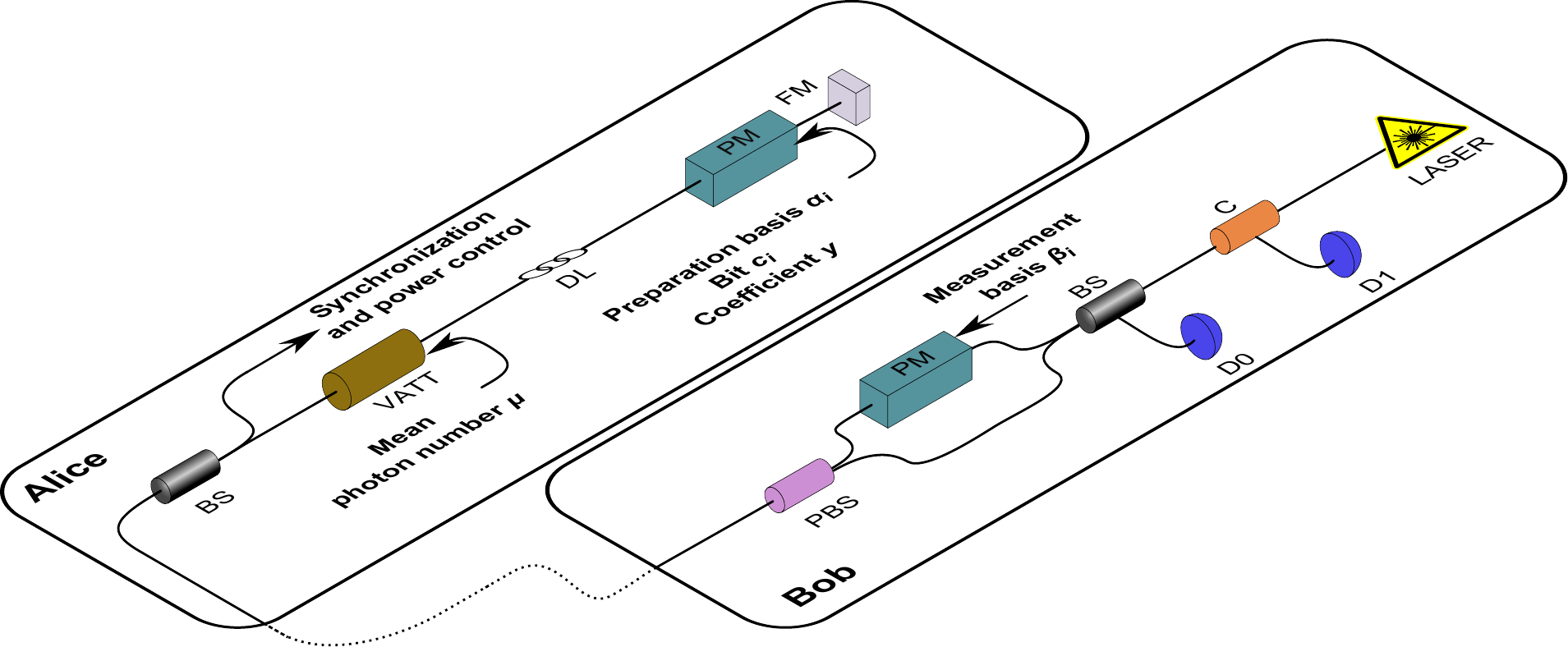}
\caption{\textbf{Experimental setup of the plug and play system.} The laser source at Bob's setup emits photon pulses at 1550 nm, which are separated at a 50/50 beamsplitter and then recombined at a polarization beam splitter, after having traveled through a short and a long arm. The latter contains a phase modulator and is appropriately arranged to transform horizontally polarized to vertically polarized states and vice versa. The pulses then travel to Alice through the communication channel, are reflected on a Faraday mirror, appropriately modulated and attenuated, and travel back to Bob orthogonally polarized. As a result, the pulses now take the other path at Bob's side and arrive simultaneously at the beamsplitter, where they interfere. Finally, they are detected by two InGaAs avalanche photodiode (APD) single-photon detectors. To implement the quantum coin flipping protocol, Alice chooses her basis and bit values by applying a suitable phase shift to the second pulse with her phase modulator. This modulator is also used to apply the state coefficient $y$. She also uses her variable attenuator to apply the required attenuation for a desired mean photon number per pulse $\mu$. Bob chooses his measurement basis by applying an appropriate phase shift at the first pulse on its way back using his phase modulator. This interferometric setup compensates for all fluctuations in the channel, which guarantees an excellent system stability. C: Circulator, BS: Beam Splitter, D0,D1: APD Detectors, PM: Phase Modulator, FM: Faraday Mirror, VATT: Variable Attenuator, PBS: Polarization Beam Splitter, DL: Delay Line.}\label{clavis2}
\end{figure*}

This analysis allows us to evaluate the maximal cheating probability attained for any given honest abort probability, for different communication distances. A security analysis allowing for a non-zero honest abort probability has also been performed for classical coin flipping protocols \cite{HW:tcc11}, providing the cheating probability bound $p_c = 1-\sqrt{H/2}$, for $H<1/2$, which is the honest abort probability region of practical interest. This can be used as a benchmark to assess quantitatively the advantage offered by the use of quantum resources for coin flipping.


\subsection{Experimental setup} We perform the demonstration of our quantum coin flipping protocol using a plug and play system, which is a considerably enhanced version of the commercial system Clavis$^2$ of IDQuantique \cite{IDQ}, designed for quantum key distribution. The experimental setup, shown in Fig. \ref{clavis2}, employs a two-way approach: light pulses at 1550 nm are sent from Bob to Alice, who uses a phase modulator to encode her information. The pulses are then reflected by a Faraday mirror and attenuated to the desired level before being sent back to Bob. Finally, Bob chooses a measurement basis with his phase modulator and registers the detection events using two high-quality single-photon detectors. When the BB84 QKD protocol \cite{BB84} is implemented by the plug and play system, the states prepared by Alice and measured by Bob correspond to the states $\ket{\Phi_{\alpha_i,c_i}}$ used in our quantum coin flipping protocol, with $y = 1/2$. Hence, the quantum transmission stage of the QKD protocol is identical to that of the coin flipping protocol with the exception that, in the latter, $y$ should be appropriately modified to guarantee the fairness of the implemented protocol. In practice, this parameter is set using the control signal that drives Alice's phase modulator, which is also used to encode Alice's basis and bit information.

Alice sets the average photon number per pulse $\mu$ with the variable attenuator shown in Fig. \ref{clavis2}, using a previously established calibration relationship. It is important to note that typical uses of the Clavis$^2$ for QKD employ significantly higher $\mu$ values. In Methods, we provide a comprehensive description of the crucial improvements made on the system to allow it to perform quantum coin flipping, which also included appropriate adjustment of the calibration and synchronization procedures. At the end of the quantum transmission part of the protocol, we derive a set of data containing the preparation and measurement basis choices of Alice and Bob, respectively, and the measurement outcomes of Bob, similarly to the raw key data obtained in quantum key distribution experiments.


\subsection{Security of the implementation} The security proof of the basic quantum coin flipping protocol provided in Ref. \cite{PCD:pra11} takes into account standard experimental imperfections (multi-photon emission, transmission loss, detector inefficiency and dark counts), which are present in our system. This analysis, however, is based on the following three important assumptions, whose validity has to be carefully examined for our implementation: (i) Honest Alice can create each state with equal probability and independently of Bob; (ii) For the first pulse $j$ that honest Bob successfully measures, his basis $\beta_j$ and bit $b$ are uniformly random and independent of Alice; (iii) For the first pulse $j$ that honest Bob successfully measures and for any state that Alice sends and any basis that Bob measures, the probabilities that Bob obtained a specific outcome are exactly proportional to the squares of the projection of Alice's state to each basis vector.

Even though these assumptions are routinely made in theoretical security proofs, they do not necessarily hold in practice unless extra caution is taken. For example, when Alice creates her states via an entangled-photon source, by measuring one half of the entangled pair and sending the other half to Bob, then the probabilities of creating each state will depend on the individual efficiencies of the detectors she uses for her measurement.
Similarly, when Bob performs a measurement using a high-efficiency detector for the outcome 0 and a lower-efficiency one for 1, then there is a bias of his outcome that Alice may use to her advantage. Last, if a different set of detectors is used for each of the two bases, again Bob's probability of successfully measuring in one basis may be much greater than in the other one. In this case, Alice can increase her cheating probability by revealing the latter basis with higher probability, thus forcing Bob to accept with higher probability.

Let us now discuss how it is possible to ensure that the aforementioned assumptions are satisfied in our implementation. Assumptions (i) and (ii), which concern Alice's choice of states and Bob's choice of measurement bases and bit $b$, respectively, can be addressed in a similar way. Alice uses a quantum random number generator to pick the basis and bit that define her state, and applies the selected phase shift using her phase modulator. Similarly, Bob uses his quantum random number generator to generate his bit $b$ and to pick the measurement basis, which is followed by the corresponding phase shift applied by his phase modulator. Therefore, in both cases, possible deviations from the uniform distribution can result from the bias of the quantum random number generators and the variation in the capability of the phase modulators to apply different phase shift values. These effects are expected to be quite small; indeed, this is confirmed by extensively analyzing the data obtained from our experimental setup and deriving appropriate bounds for these deviations (see Supplementary Note 1 for details).

Assessing assumption (iii), which concerns Bob's basis and outcome distributions given a detection event, turns out to be particularly important. We would like to ensure that when Bob has registered a detection this happens on each one of the two bases with uniform probability. In our experimental setup, the same set of detectors is used for both bases so we do not expect an important deviation from a uniform distribution, which is again confirmed by our data. However, we observe a significant asymmetry in the number of detections registered by each detector; in fact, the ratio of the detection efficiencies is found to be approximately 0.68 (see Supplementary Table 1). This asymmetry can clearly be used by malicious Alice to increase her cheating probability. To remedy this problem, we implement a simple yet powerful solution proposed in Ref. \cite{NJM:natcom12}, the symmetrization of losses. Bob effectively makes the detection efficiencies equal by throwing away some detection events from the detector featuring higher efficiency. Even after this symmetrization procedure, an uncertainty on the efficiency ratio remains and can be appropriately bounded.

In Supplementary Note 1, we provide a rigorous security analysis of the basic quantum coin flipping protocol, when it is implemented using a practical plug and play system. This analysis takes into account all standard imperfections, as well as the additional inevitable imperfections present in our system that were previously discussed, and is used to derive the cheating probabilities that are necessary to assess quantitatively the advantage offered by quantum communication for coin flipping.


\begin{center}
\begin{table*}
\begin{tabular}{|c|c|c|c|c|}
\cline{1-5}
         &\multicolumn{2}{c|}{15 km} & \multicolumn{2}{c|}{25 km}\\
\cline{1-5}
Coefficient $y$ &\multicolumn{2}{c|}{0.88} &  \multicolumn{2}{c|}{0.85}\\
\cline{1-5}
$\mu$  ($\times 10^{-3})$            &   2.8 $\pm$ 0.1       &  {\bf 2 $\pm$ 0.1}  & 5 $\pm$ 0.1   & {\bf 4 $\pm$ 0.1}\\
\cline{1-5}
Protocol rounds $K$                                    &     88000                   & {\bf 131000} &   130000 &  {\bf 174000}\\
\cline{1-5}
~~~~~Cheating probability~~~                     & ~~~0.916 $\pm$ 0.002~~~ &~ ~{\bf 0.914 $\pm$ 0.002}  ~ &~~ 0.947 $\pm$ 0.003 ~  &~  {\bf 0.942 $\pm$ 0.003}~\\
\cline{1-5}
\end{tabular}
\caption{\textbf{Experimental parameter values for $\bm{H = 0.8\%}$.} The parameters correspond to a fair protocol, which is ensured by the choice of the coefficient $y$. The uncertainty in the values of $\mu$ is due to the difference between the value expected from Alice's calibrated variable attenuator setting and the value deduced from Bob's detection events together with the known losses in the path between Alice and Bob. The number of detection events used to calculate $K$ is sufficiently large (typically $10^6$) to ensure negligible finite-size effects in our implementation; for instance, for $H = 0.8\%$, the probability that the honest abort probability is greater by more than $0.2\%$ is $10^{-9}$. The cheating probability is computed using the security analysis of the basic quantum coin flipping protocol for the plug and play implementation. The numbers in bold correspond to the values shown in Fig. \ref{fig:ha_cheat}.}
  \label{table}
\end{table*}
\end{center}

\vspace{-1cm}
\subsection{Experimental quantum coin flipping results} We perform quantum coin flipping experiments for two channel lengths, namely 15 and 25 km. In Table \ref{table} we provide typical values of the experimental parameters used in the implementations. Based on the data obtained from the quantum transmission part of the protocol and taking into account the symmetrization procedure, we calculate the number of protocol rounds $K$ that are required to achieve a desired honest abort probability. The detection events registered by Bob, in conjunction with the known experimental conditions in the path between Alice and Bob, can be used to determine the actual average photon number per pulse $\mu$ that is exiting Alice's system. In practice, we find that this value is slightly different from the one estimated by the variable attenuator calibration relationship. This difference is at the origin of the uncertainty in the values of $\mu$ shown in Table \ref{table}.

This procedure is performed using several values of $\mu$ for each channel length, and then choosing the number of rounds $K$ to attain the desired honest abort probability. Based on these sets of parameters, we derive the cheating probabilities of a malicious Alice and Bob, $p^A_{q}$ and $p^B_{q}$, respectively, using the extended security analysis of the basic quantum coin flipping protocol (see Supplementary Note 1 for the full expressions). This allows us to find, for both channel lengths, the sets of values for $\mu$, $K$ and $y$ that minimize the cheating probability and at the same time make the protocol fair ($p^A_{q} = p^B_{q} \equiv p_q$). Note that for simplicity, the $y$ values of our experimental data have been chosen independently of the honest abort probability value; in practice, slight modifications of these values might be required to achieve a perfectly fair protocol for each specific honest abort probability. The optimized experimental parameters for an honest abort probability $H = 0.8\%$ are shown in bold in Table \ref{table}.

\begin{figure}
\includegraphics[width=8.5cm]{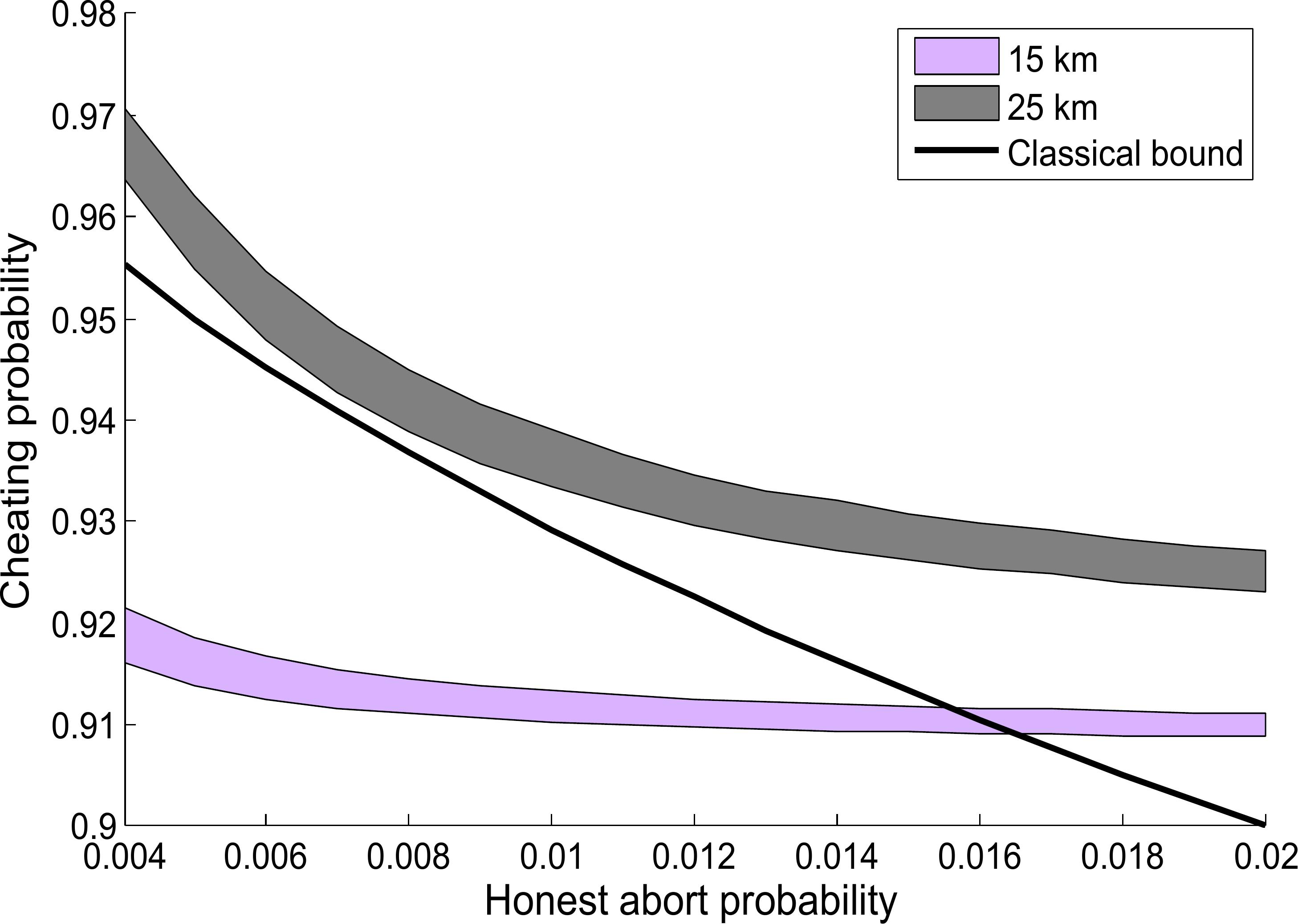}
\caption{\textbf{Cheating vs honest abort for 15 and 25 km.} The cheating probability for each honest abort probability value is calculated from the experimental data using the security analysis of the basic quantum coin flipping protocol adapted to our implementation. The values correspond to a fair protocol. The shaded areas are derived from the uncertainty in the estimation of the average photon number per pulse exiting Alice's setup. The solid line represents the cheating probability bound for classical coin flipping protocols. For a 15 km channel length, quantum communication leads to lower cheating probability values than is classically possible, for a wide range of practical honest abort probabilities. The cheating probabilities derived at 25 km are always greater than the classical bound.}
\label{fig:ha_cheat}
\end{figure}

In Fig. \ref{fig:ha_cheat} we show the cheating probability calculated from our experimental data for 15 and 25 km, as a function of the honest abort probability. For each value of the honest abort probability, the number of rounds $K$ and mean photon number per pulse $\mu$ has been optimized as explained previously. The uncertainty in the estimation of $\mu$ is illustrated by the shaded areas in the plot. To quantify the advantage offered by quantum communication, we use the classical cheating probability bound, $p_c$ \cite{HW:tcc11}. We can see that the cheating probability is strictly lower than classically possible for a distance of 15 km, for a wide range of practical values of the honest abort probability.
In particular, taking into account finite-size effects, for a range of honest abort probability from $0.4\%$ to $1.45\%$,  the cheating probability is lower than classically possible unless with probability of the order $10^{-9}$.
The area corresponding to the data obtained at 25 km is just above the classical cheating bound for all honest abort probability values, which means that a quantum advantage cannot be claimed in this case.

\begin{figure}
\includegraphics[width=8.5cm]{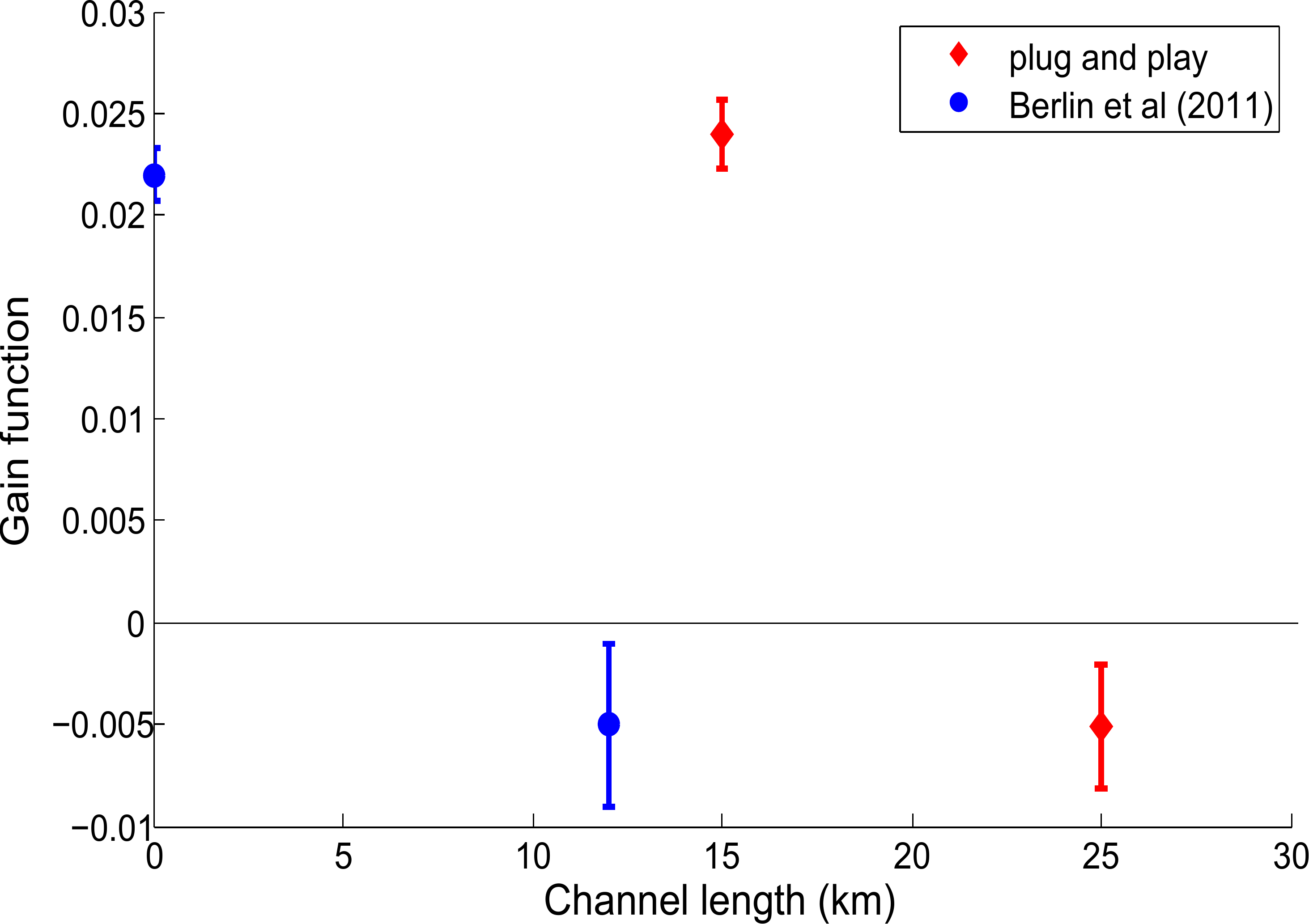}
\caption{\textbf{Gain as a function of channel length.} The gain function is calculated as the difference between the classical and quantum cheating probability, and illustrates the advantage offered by the use of quantum communication for coin flipping. The error bars are computed from the uncertainty in estimating the average photon number per pulse in Alice's setup. The diamonds correspond to the cheating probabilities achieved by our plug and play implementation for 15 and 25 km, for a fixed honest abort probability of 0.8\%. A positive gain is obtained for 15 km, while the gain remains below the classical limit for 25 km. For comparison, previous experimental results based on an entangled-photon source implementation of a loss-tolerant quantum coin flipping protocol \cite{BBB:natcomm11} are also shown (circles): a positive gain was experimentally obtained for a distance of 10 m, with an honest abort probability of 1.8\%, while no positive gain was possible for a distance of 12 km.}\label{fig:gain}
\end{figure}

To obtain further insight into our results, we define a gain function, as follows:
\[G = p_c-p_q,\]
where $p_c$ and $p_q$ are the classical cheating probability bound and the quantum cheating probability value derived from our experimental data, respectively. If the experimental data yields a positive $G$ for a certain honest abort probability, this means that these results cannot be obtained by any purely classical means. We can then use the gain as a figure of merit to assess the performance of our quantum coin flipping implementation in a secure communication scenario. In Fig. \ref{fig:gain}, we show the gain as a function of distance, for a fixed honest abort probability $H = 0.8\%$. For the channel length of 15 km, a distance which is sufficient for many applications requiring communication over metropolitan area networks, the gain is of the order of $0.025$, while for the channel length of 25 km no positive gain can be obtained.

Note that in the distrustful model with information-theoretic security, it is not known if it is possible for Alice and Bob to collaborate in order to increase the robustness of the implementation. This results in an inherent limitation to the attainable communication distance in our quantum coin flipping implementation. However, using better single-photon detectors with lower dark count rates \cite{NTH:sst12} for instance, can readily extend the range of our protocol.

Finally, in our implementation, the classical steps of the coin flipping protocol following the quantum transmission are not performed in real time. However, it is clear that the coin flipping rate will be dominated by the time that it takes for $K$ pulses to travel from Alice to Bob. For a laser pulse repetition rate of 10 MHz, this corresponds roughly to a few tens of coin flips per second. As we can see in Table \ref{table}, if Alice increases the average photon per pulse exiting her system, the required number of protocol rounds reduces, which also reduces the runtime for the protocol, but this comes at the expense of a slightly higher cheating probability. Again, using better single-photon detectors can result in a substantially lower number of required rounds. In a real communication scenario of two distrustful parties wishing to agree on a coin value using the plug and play system, the parties would be given a choice of gain values for a range of honest abort probabilities given their communication distance and the desired communication rate.

\subsection{Enhancing security against bounded adversaries} We have seen that our basic quantum coin flipping protocol achieves information-theoretic security, which is impossible classically. However, this security level comes at a price of a high bias; indeed, as we see in Fig. \ref{fig:ha_cheat}, the unbounded adversary can bias the coin with probability greater than 90\%. This might not be suitable for some applications. It is then interesting to consider combining our protocol with protocols that achieve a bias asymptotically close to zero against bounded adversaries. Combining protocols with different types of security is in fact a powerful concept, which is widely used in practice. This allows communications to remain secure not only at the present time but also in the future, accommodating at the same time for different types of adversaries with unknown or rapidly evolving technological and computational capabilities.

To construct combined protocols for quantum coin flipping, we apply the following general lines: We discern three stages, as in the commonly used protocols against bounded adversaries, including classical protocols employing one-way functions \cite{goldreich01} and quantum protocols in the noisy quantum storage model \cite{WST:prl08,NJM:natcom12}. In the first stage (commit), which remains unchanged from the protocols against bounded adversaries, Alice and Bob exchange classical or quantum messages such that at the end of this stage each party has almost perfectly committed to one bit, $S$ and $T$, respectively. In the second stage (encrypt), Alice and Bob encrypt their respective random bits using the committed values. In particular, Alice sends $K$ pulses using the states $\ket{\Phi_{\alpha_i,c_i \oplus S}}$, for $i=1,...,K$, and Bob replies by sending $T\oplus b$ as well as $j$, the index of the first measured pulse, as in the basic protocol. In the third stage (reveal), Alice and Bob reveal $(c_j, S)$ and $(b, T)$, respectively, together with additional information depending on the underlying bounded adversary model and, if nobody aborts, the value of the coin is $c_j \oplus b$ (see Supplementary Note 2 for explicit combined protocols for models with computationally bounded adversaries and adversaries with noisy quantum storage).

The combined protocols constructed as explained above achieve an almost perfect security against bounded adversaries, exactly as the original protocols; in addition, when the adversaries are unbounded, they still cannot cheat with a probability higher than the one provided by our basic quantum coin flipping protocol, which is strictly better than classically possible. Hence, these protocols offer the maximal possible security guarantees.


\section{Discussion}
\noindent The results that we have presented constitute one of the few instances of a rigorously proven and demonstrated advantage of quantum over classical communication, which can be used for practical secure communications between distrustful parties with information-theoretic security guarantees. We have demonstrated this advantage using a practical plug and play system over distances suitable for metropolitan area communication networks. This enlarges the scope of quantum cryptography, in particular to practical applications where the parties do not trust each other.

We emphasize that dealing with distrustful parties is more complicated than the quantum key distribution scenario, both in theory and in practice. For example, although randomized procedures like error correction and privacy amplification that are widely employed in QKD have been used in the security analysis of protocols dealing with bounded adversaries \cite{WCS:pra10}, it is an open question whether such procedures can be used in the information-theoretic security setting; in principle, any such step can be used by the malicious party to his or her advantage. Therefore, new techniques may be needed in order to deal with the imperfections of the implementation and the inherent limitations to the attainable communication distance. Our results bring quantum cryptography in the distrustful model at a comparable level of practicality as quantum key distribution and provide means to benchmark this type of primitives in a way similar to QKD protocols.

Additionally, by combining our quantum coin flipping protocol with protocols secure against bounded adversaries we enhance those with a level of information-theoretic security. This assures that an honest party will always obtain security guarantees stronger than possible by classical means. It is also interesting to note that our protocol is based on a bit commitment scheme, augmented only by an additional classical message from Bob to Alice between the commit and reveal stages. This means that our combined coin flipping protocols can also be viewed as commitment schemes where both parties commit some value to each other. Hence, our security analysis can be extended in a straightforward way to hold for bit commitment in the computational models that we have considered. In the same way, our implementation indeed performs plug and play quantum bit commitment. We also note that a weaker, but still very powerful, variant, called weak coin flipping, with almost perfect information-theoretic security is in theory possible with quantum technology \cite{Moc:focs04,acgkm:arxiv14}. Our implementation is a first step towards making such protocols a reality, however the quantum protocols that achieve almost perfect security are not well understood and currently necessitate large-dimension entangled states. Simplifying such protocols is an important open question.

Last, as in practical quantum key distribution, our implementation of quantum coin flipping may be vulnerable to side-channel attacks (see Methods for details). The power control setup placed at the entrance of Alice's system as a countermeasure for the so-called Trojan horse attacks \cite{GFK:pra06} can also be used by Alice to properly characterize the photon distribution of the pulses sent by Bob \cite{ZQL:njp10}. This is important to counter, for instance, an attack by which Bob sends strong light pulses to Alice, which lead to an increased average photon number per pulse and consequently to a greater cheating probability. Identifying potential side-channel attacks and devising appropriate countermeasures is of great importance, as for all practical quantum cryptographic systems.


\section{Methods}
\subsection{Plug and play quantum coin flipping system}
Our quantum coin flipping implementation is based on the commercial quantum key distribution system Clavis$^2$ of IDQuantique. Using a QKD system for an implementation of a cryptographic primitive that requires an entirely different security analysis and operates in non-standard experimental conditions necessitated several important modifications to the system. First, single-photon detectors with very low dark count rates were installed in the quantum coin flipping system; indeed, the honest abort probability is very sensitive to this parameter and so with even moderately high dark counts the quantum advantage vanishes at any distance. The dark count rate per detection gate of the detectors $D_0$ and $D_1$ were $7\times 10^{-6}$ and $1.6\times 10^{-6}$, with corresponding quantum efficiency values of 7.7\% and 5.2\%, respectively. Second, new functionalities and control signals were added to the system to be able to apply the coin flipping protocol, in particular, those allowing us to rotate the standard BB84 states so that the optimal states for a fair protocol could be used and those allowing us to reduce the mean photon number per pulse $\mu$ at suitable values for coin flipping. These values were in fact one or two orders of magnitude lower than those typically used for QKD. This last point was also crucial for many aspects of the implementation, since a very low $\mu$ value hindered the operation of several embedded calibration and testing processes of the system, which were therefore entirely redesigned. Such calibration procedures play an important role in the two-way configuration of the plug and play system, which imposes particular care in the synchronization of the phase shift and attenuation signals, and the detection gates. Among those, of particular importance is the calibration procedure involving the variable attenuator at Alice's site, which was actually the main source of the uncertainty that we observe in our data. Finally, the QKD classical post-processing procedures were replaced by our software, which used as an input the raw data of quantum signal exchange between Alice and Bob, together with basis choice information. These enhancements led to the development of a practical, plug and play system that is capable of performing quantum coin flipping in addition to key distribution.

It is important to note that the advantage of the plug and play system with respect to other systems providing the functionalities required by our protocol is that it offers a particularly robust and stable implementation, which allows to perform experiments at low signal level for long time duration, resulting in very reliable results. This system can also potentially be used for protocols employing decoy states \cite{LMC:prl05,Wan:prl05,ZQM:prl06}. Although the use of decoy states is a powerful tool for achieving practical long-distance quantum key distribution and for improving the performance of quantum cryptographic protocols in the noisy storage model \cite{WCS:pra10}, it is not known, to the best of our knowledge, if a protocol employing decoy states can be devised for quantum coin flipping providing security against all-powerful adversaries.


\subsection{Side channels and practical security} Our quantum coin flipping implementation takes explicitly into account the standard imperfections (multi-photon emission, transmission loss, detector inefficiency and dark counts) present in practical systems. We also consider imperfections related to asymmetries in the detection efficiency and basis-dependent flaws in the components of Alice's and Bob's devices, which play a crucial role for the practical security of the implementation. It is clear, however, that similarly to QKD experiments, further deviations between the security proof and the actual implementations inevitably exist and can lead to side-channel attacks by the adversary. Although an exhaustive analysis of possible side channels is out of the scope of the present work, we examine a few prominent cases known in the context of QKD demonstrations, some of which are particularly relevant for the plug and play system at the basis of our implementation.

In the setting of quantum key distribution, an efficient eavesdropping attack that also applies to the plug and play system consists in shifting in time the second pulse (see Fig. \ref{clavis2}) such that this pulse is only partially modulated by Alice's phase modulator. This so-called phase remapping attack \cite{FQT:pra07,XQL:njp10} effectively alters the relative phase between the two pulses and allows Eve to obtain key information for ranges of quantum bit error rate values that would otherwise be considered acceptable. In the distrustful setting of quantum coin flipping, malicious Bob attempts to maximize his cheating probability by performing an optimal measurement to the received states, and so reducing the probability of distinguishing them cannot help him. Additionally, errors in the state preparation performed by Alice have been considered in detail in the security analysis of our implementation (see Supplementary Note 1). Similarly, attacks exploiting the loophole introduced by detection efficiency mismatch, such as the time-shift attack \cite{MAS:pra06,QFL:qic07,ZFQ:pra08}, are, in principle, excluded by the symmetrization procedure included in our experimental protocol. Finally, an effective countermeasure against the powerful blinding attack \cite{LWW:natphoton10}, where the single-photon detectors are brought to a classical operation regime and can be fully controlled by the adversary, consists in randomly suppressing detector gates and emitting an alarm signal in case of registered detection events during those gates \cite{patent}. This countermeasure is implemented in our system.

In addition to the aforementioned side-channel attacks, it is important to note the issue of phase randomization \cite{ZQM:apl07}, which is an assumption typically made in security proofs and hence should be satisfied in practice. Phase randomization together with suitable intensity monitoring are also required for the characterization of an untrusted source, which is particularly relevant for the plug and play system \cite{ZQL:njp10}. Although all the hardware components necessary for implementing active phase randomization and source characterization at Alice's site are available in our system, these processes were not performed in real time, due mainly to the difficulty in generating random real numbers in real time and to the limited bandwidth of the threshold discriminator used for intensity monitoring. Clearly, for any real-life implementation, following such procedures is essential.


\section{Acknowledgments}

\noindent This research was supported by the French National Research Agency, through CRYQ (ANR-09-JCJC-0067) and HIPERCOM (2011-CHRI-006) projects, by the European Union through the project Q-CERT (FP7-PEOPLE-2009-IAPP) and the ERC project QCC, and by the City of Paris through the CiQWii project. ID Quantique work was supported by the European Union project SIQS. A.P. and T.L. acknowledge support from Digiteo. P.J. acknowledges support from the ANRT (Agence Nationale de la Recherche et de la Technologie). A.P., P.J., T.L., and E.D. thank the ID Quantique team for their hospitality during their visits to Geneva.

\clearpage

\section{Supplementary Figures}
\begin{supfigure}[h!]
\includegraphics[width=2.5in]{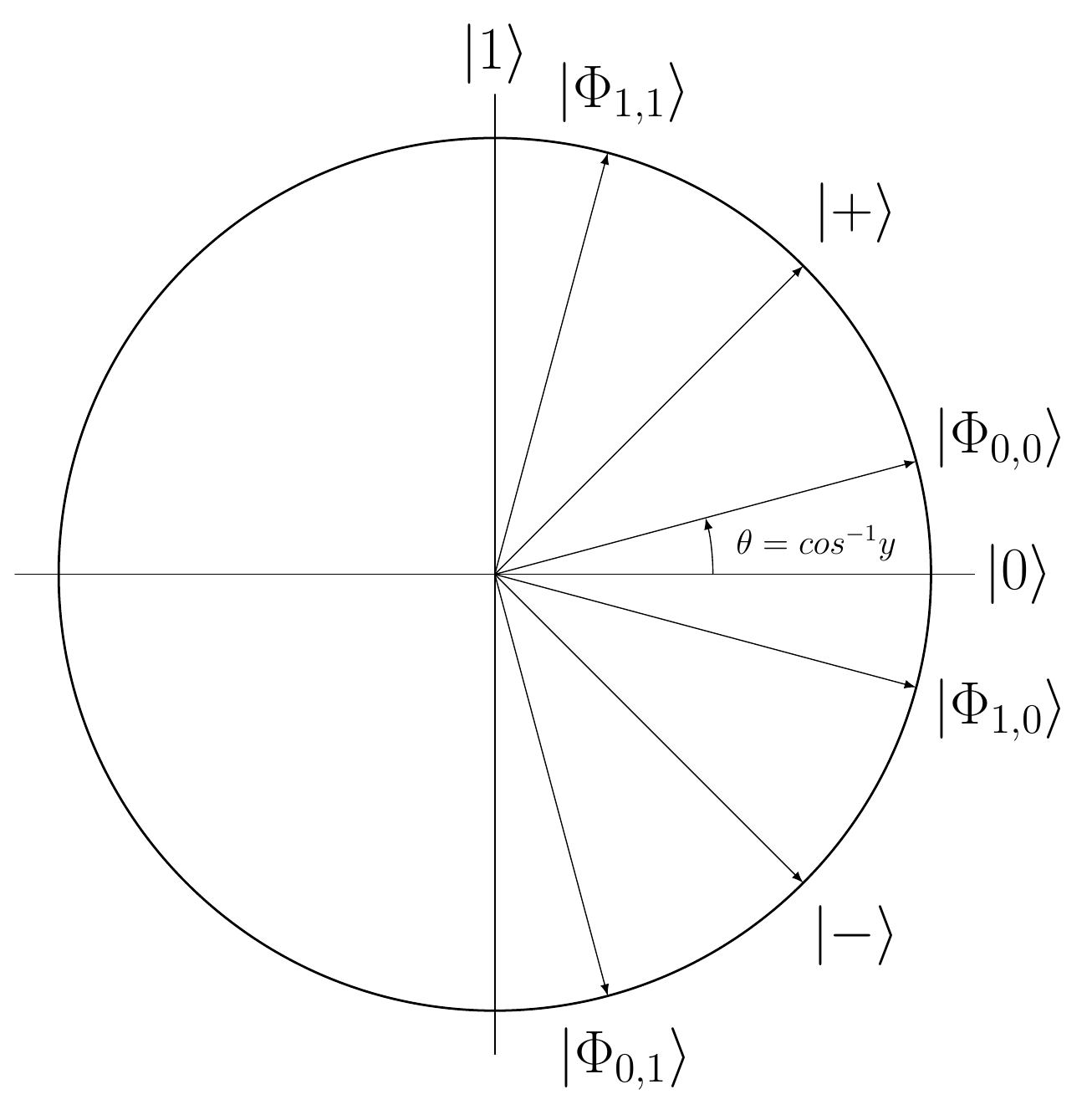}
\caption{\textbf{Protocol States.} For choice of basis $\alpha\in\{0,1\}$ and bit $c\in\{0,1\}$, the honest states of the protocol are of the form $|\Phi_{\alpha,c}\rangle$, where $\ket{\Phi_{\alpha,0}}=\sqrt{y}\lvert0\rangle+(-1)^{\alpha}\sqrt{1-y}\lvert1\rangle$ and $\lvert\Phi_{\alpha,1}\rangle=\sqrt{1-y}\lvert0\rangle-(-1)^{\alpha}\sqrt{y}\lvert1\rangle$, with $y \in (\frac{1}{2},1)$.} \label{fig:states}
\end{supfigure}

\section{Supplementary Tables}

\begin{center}
\begin{suptable}
\begin{center}
\begin{tabular}{|c|c|c|c|}
\cline{1-4}
Basis & Bit & Detections (15 km) & Detections (25 km)\\
\cline{1-4}
0 & 0 & 84071 & 54915\\
\cline{1-4}
0 & 1 & 53200 & 34994\\
\cline{1-4}
1 & 0 & 82825 & 54279\\
\cline{1-4}
1 & 1 & 51497 & 34252\\
\cline{1-4}
\end{tabular}\end{center}\caption{\textbf{Detection events for 15 and 25 km.} These results correspond to the case when the preparation basis and bit values of Alice agree with the measurement basis and outcome of Bob respectively. For the 15 km experiment, the total number of pulses that were sent was $1.1458\times10^{10}$ and the total number of detection events is 593272. For the 25 km experiment, $7.2905\times10^{9}$ pulses were sent and 414649 detection events were registered in total.}
\label{table1}
\end{suptable}
\end{center}



\vspace{3in}
\section{Supplementary Notes}
\subsection{Supplementary Note 1 - Security Analysis}

Here we provide the security analysis of the basic quantum coin flipping protocol (see Box 1 in main text), when it is implemented using a practical plug and play system. This analysis is an extension of the security proof provided in Ref. \cite{PCD:pra11} and takes into account standard experimental imperfections (multi-photon emission, transmission loss, detector inefficiency and dark counts), as well as additional imperfections specific to our experimental setup that we explain below.

We start by stating the security assumptions made in the security proof of Ref. \cite{PCD:pra11} and provide the security analysis when these assumptions are satisfied. Then, we show how to ensure that these assumptions are almost satisfied by our implementation. Last, we investigate how the inevitable small deviations of our implementation from these assumptions affect the cheating probabilities and conclude that there is a clear quantum advantage for the 15 km channel length implementation.

\subsubsection{Security assumptions and analysis with standard imperfections.} We first provide a security analysis when the following three assumptions are satisfied:
\begin{enumerate}
\item Honest Alice creates each of the four protocol states $\ket{\Phi_{\alpha_i,c_i}}$ (see Suppl. Fig. \ref{fig:states}) with the same probability, independently for each pulse and independently of Bob.
\item For the first pulse $j$ that honest Bob successfully measures, the distribution of his measurement basis $\beta_j$ and his bit $b$ is uniform and independent of Alice.
\item For the first pulse $j$ that honest Bob successfully measures, if the state of the pulse is $\rho$, then for each basis $\beta_j$, the probabilities of the two outcomes are $ \bra{\Phi_{\beta_j,0}} \rho \ket{\Phi_{\beta_j,0}}$ and $ \bra{\Phi_{\beta_j,1}} \rho \ket{\Phi_{\beta_j,1}}$.
\end{enumerate}

We describe the optimal cheating strategies of malicious Alice and Bob and derive the corresponding cheating probabilities when the above assumptions are satisfied.\\

\paragraph{Malicious Alice.} Let us assume that Alice tries to bias the coin towards the value $x=0$ (the analysis for $x=1$ is similar). We also assume that Bob successfully measured the first pulse, thus providing an upper bound to Alice's cheating probability. Honest Bob has therefore picked a uniformly random basis $\beta$ and has detected the qubit sent by Alice. He replies with the uniformly random bit $b$ in the next step of the protocol. Alice then has to reveal a basis $\alpha$ and the value $c=b$, so that the coin value is 0. If $\alpha \neq \beta$, Bob accepts. If $\alpha=\beta$, Bob checks whether his measurement outcome agrees with $c$. Note that here and in the following, we drop for simplicity the index $j$ used in the description of the basic quantum coin flipping protocol for the first measured pulse by Bob, which is used to establish the coin value.

The analysis is exactly the same as in Ref.  \cite{BBBG:pra09}, which is based on the rigorous analysis in Ref. \cite{SR:QIC02} of the original protocol of Ref. \cite{ATVY:stoc00}. The only difference between our protocol and the one in Ref. \cite{ATVY:stoc00}, is that, there, Bob waits for Alice's announcement of the basis before measuring. Hence, the analysis in Ref. \cite{SR:QIC02} corresponds to the case that Alice reveals the same basis as Bob; in the other case, she can perfectly cheat since Bob always accepts.

Alice's optimal strategy consists of finding the state that will maximize the average probability of revealing bit 0 or 1 (since Bob's choice $b$ is uniform). However, when Alice needs to reveal $c=0$ or $c=1$, she also has the choice of which basis to reveal. This might enable Alice to increase her cheating probability by creating a state in a large Hilbert space, sending part of it to Bob and after Bob's message, performing some general operation on her part to decide which basis to reveal. Nevertheless, this analysis has already been done in Ref. \cite{SR:QIC02}. More precisely, in Ref. \cite{SR:QIC02} (section 6.4.1), it is first shown that for any protocol where there are two honest pure states that correspond to bit 0 and two honest pure states that correspond to bit 1 (as in our case), Alice's optimal strategy is to send the state that maximizes the average probability of revealing  $(\alpha=0,c=0)$ and $(\alpha=1,c=1)$ or of revealing $(\alpha=1,c=0)$ and $(\alpha=0,c=1)$. In high level, this is true since the states in the pairs $\{ \ket{\Phi_{0,0}}, \ket{\Phi_{1,1}}\}$ and $\{ \ket{\Phi_{1,0}}, \ket{\Phi_{0,1}}\}$ are closer to each other than the orthogonal states in the pairs $\{ \ket{\Phi_{0,0}}, \ket{\Phi_{0,1}}\}$ and $\{ \ket{\Phi_{1,0}}, \ket{\Phi_{1,1}}\}$  (see Suppl. Fig. \ref{fig:states}).

For the first case, a simple calculation (see Ref. \cite{SR:QIC02}) shows that the optimal over all possible states is in fact the pure state $\ket{+}$; then, after reception of Bob's bit $b$, Alice reveals $(\alpha=b,c=b)$. For the second case, the optimal state is shown to be the pure state $\ket{-}$; then, Alice reveals $(\alpha=1-b,c=b)$. The probability that she forces the outcome 0 is then in both cases:
\begin{equation*}
\Pr[x=0|\text{same bases}]=\frac{1}{2}+\sqrt{y(1-y)},
\end{equation*}
where $y$ is the coefficient of the honest states. Note that Alice could also decide to prepare any mixture of the states $\ket{+}$ and $\ket{-}$ and achieve exactly the same cheating probability. When the bases are different, according to the protocol, Bob always accepts the coin. Since Bob's basis choice is uniformly random and independent of Alice, Alice can bias the coin with probability:
\begin{equation}
p_q^A \leq \frac{3}{4}+\frac{1}{2}\sqrt{y(1-y)}
\label{eq:pqA}\vspace{0.3in}
\end{equation}

\paragraph{Malicious Bob.}
The optimal cheating strategy of an all-powerful Bob is complex and involves his ability to know the number of photons in each of the $K$ pulses sent by Alice. He can then accordingly optimize his POVM on all $K$ pulses to maximize his cheating probability. It is important to note that, under Assumption 1, honest Alice uses a uniformly random bit $c_i$ to prepare the state in each pulse $i$, and all $c_i$s are independent of each other. 
We upper bound Bob's cheating probability by considering that his cheating probability is 1 in all cases except for four events $A_i$ ($i=1,\dots,4$), for which we find appropriate bounds as shown below.

Let us assume, without loss of generality, that Bob's desired outcome is $x=0$ and let $\Pr[x=0|A_i]$ be the probability that Bob will force his preference when event $A_i$ has taken place, which happens with probability $\Pr[A_i]$. According to the protocol, the number of photons per pulse $i$ follows the Poisson distribution $p_i=\mu^i e^{-\mu}/i!$, where $\mu$ is the mean photon number. We consider the following events, for $K$ number of rounds:
\begin{description}
\item[$A_1$]: Bob receives only vacuum pulses. This event occurs with probability $\Pr[A_1]=e^{-\mu K}$. Since Bob has no knowledge of Alice's bit, which is uniformly random, he picks a random bit, and hence $\Pr[x=0|A_1]=1/2$.
\item[$A_2$]: Bob receives vacuum pulses, at least one single-photon pulse and no two- or more-photon pulses. This event occurs with probability $\Pr[A_2]=(p_0+p_1)^K-p_0^K$. We will assume here that Bob does not actually receive any vacuum pulses, which can only increase his cheating.

To analyze cheating Bob, we use the very strong loss-tolerant properties of our coin flipping protocol.
From the definition of the states of the honest protocol (with $y \in (\frac{1}{2},1)$), we have
\begin{eqnarray*}
\rho_0&=&\frac{1}{2}\ketbra{\Phi_{0,0}}{\Phi_{0,0}} +\frac{1}{2}\ketbra{\Phi_{1,0}}{\Phi_{1,0}}\\
\rho_1&=&\frac{1}{2}\ketbra{\Phi_{0,1}}{\Phi_{0,1}}  + \frac{1}{2}\ketbra{\Phi_{1,1}}{\Phi_{1,1}}
\end{eqnarray*}
 It is easy to see that the maximum eigenvalue of $\rho_0$ and $\rho_1$ is equal to $y$ and their minimum eigenvalue is equal to $(1-y)$ (in fact, $\rho_0 = y \altketbra{0} + (1-y) \altketbra{1}$ and $\rho_1 = (1-y) \altketbra{0} + y \altketbra{1}$). Hence,
\[ \forall m \in \{0,1\},\;\;\; 2(1-y) \mathbb{I} \preceq \rho_{m} \preceq 2y\mathbb{I},
\]
where $\mathbb{I}$ is the totally mixed state and $A \preceq B$ means that the matrix $B-A$ is positive.  From the above, we can conclude that
\[
 \rho_{0} \succeq \frac{1-y}{y} \rho_1 \;\;\; , \;\;\;  \rho_{1} \succeq \frac{1-y}{y} \rho_0.
\]
Hence, there exist positive norm-1 matrices $\xi_0$ and $\xi_1$ such that
\begin{eqnarray*}
 \qquad   \;\;\rho_{0} = \frac{1-y}{y} \rho_1 + \frac{2y-1}{y}\xi_0 ,\;
 \rho_{1} = \frac{1-y}{y} \rho_0 + \frac{2y-1}{y}\xi_1.
\end{eqnarray*}
Then, we can rewrite $\rho_0$ and $\rho_1$ as
\begin{eqnarray}\label{xi}
 \; \rho_{0} = {y} \xi_0 + (1-y)\xi_1 \; , \;
 \rho_{1} = (1-y) \xi_0 + {y}\xi_1.
\end{eqnarray}
Let $\{M_{i,b}\}_{i \in [K], b \in \{0,1\}}$ the POVM that Bob applies on all $K$ pulses to determine the index $i$ that he will announce as his first measured pulse, as well as his guess $b$ for Alice's bit $c_i$. We have $\sum_{i,b} M_{i,b} = I$. Let $M_i = M_{i,0} + M_{i,1}$ be the POVM element that corresponds to the event that Bob outputs $i$ as his first measured pulse. We have for Bob's cheating probability
\begin{align*}
\Pr[x=0|A_2] & = \sum_{i \in [K]} \Pr[\mbox{Bob outputs }(i,b = c_i)] \\
& = \sum_i \Pr[i] \cdot \Pr[ b = c_i |  i]
\end{align*}
Fix any index of a pulse $j$ and a value $c_j$ for Alice's bit on pulse $j$. In this case, Bob's density matrix is $\rho_{c_j}$ in the register that corresponds to the $j$-th pulse and the totally mixed state $\mathbb{I}$ in the other registers. We denote this state by $\rho_{c_j} \otimes \mathbb{I}_{\neg{j}}$. Note that the necessary  property is that this is a product state and the state that corresponds to the remaining pulses is independent of $c_j$. The proof below would also work for any other state instead of $\mathbb{I}_{\neg{j}}$, given that this state is also independent of $c_j$.

When Bob performs the above POVM and outputs index $j$, then after the measurement he has the (unnormalized) mixed state $M_j(\rho_{c_j} \otimes \mathbb{I}_{\neg{j}})$. Our goal is to determine how well Bob can guess the value $c_j$ when he outputs the index $j$, in other words, how well he can distinguish the states $M_j(\rho_{0} \otimes \mathbb{I}_{\neg{j}})$ and $M_j(\rho_{1} \otimes \mathbb{I}_{\neg{j}})$. Using the optimality of the Helstrom measurement, Eq. (\ref{xi}), and the fact that $\rho_0 + \rho_1 = \xi_0 + \xi_1$, we have that:
\begin{align*}
\qquad\lefteqn{\Pr[b = c_j |  j]} \\
&\le \frac{1}{2} + \frac{1}{2}\cdot \frac{||M_j(\rho_{0} \otimes \mathbb{I}_{\neg{j}}) - M_j(\rho_{1} \otimes \mathbb{I}_{\neg{j}})||_{tr}}{Tr(M_j(\rho_{0} \otimes \mathbb{I}_{\neg{j}})) + Tr(M_j(\rho_{1} \otimes \mathbb{I}_{\neg{j}}))}\\
& = \frac{1}{2} + \frac{1}{2} \cdot \frac{(2y-1)||M_j(\xi_{0} \otimes \mathbb{I}_{\neg{j}}) - M_j(\xi_{1} \otimes \mathbb{I}_{\neg{j}})||_{tr}}{Tr(M_j(\xi_{0} \otimes \mathbb{I}_{\neg{j}})) + Tr(M_j(\xi_{1} \otimes \mathbb{I}_{\neg{j}}))}\\
& \leq y
\end{align*}
Since the above holds for any $j$, we can conclude that
\begin{align*}
\Pr[x=0|A_2] = \sum_i \Pr[i] \cdot \Pr[b = c_i |  i] \le y
\end{align*}

\item[$A_3$]: Bob receives only vacuum pulses and one two-photon pulse. This events occurs with probability $\Pr[A_3]=K p_2 p_0^{K-1}$. Let $\sigma_0$ and $\sigma_1$ the mixed states that correspond to the two-photon pulse Bob receives when Alice's bit is 0 or 1; then, the optimal measurement to distinguish these states is given by the Helstrom measurement and yields $\Pr[x=0|A_3]=\frac{1}{2}+\frac{1}{2}|| \frac{1}{2}\sigma_0 - \frac{1}{2}\sigma_1 || =  y$.

\item[$A_4$]: Bob receives vacuum pulses, one two-photon pulse and at least one single-photon pulse. This events occurs with probability $\Pr[A_4]=Kp_2[(p_0+p_1)^{K-1}-p_0^{K-1}]$. As before, we assume that Bob receives no vacuum pulses, which can only increase his cheating.

Let $\{M_{i,b}\}_{i \in [K], b \in \{0,1\}}$ the POVM Bob applies on all $K$ pulses to determine the index $i$ he announces as his first measured pulse, as well as his guess $b$ for Alice's bit $c_i$. Let $j$ the index that corresponds to the two-photon pulse. We have
\begin{align*}
\Pr[x=0|A_4] = \sum_{i \in [K]} \Pr[\mbox{Bob outputs} (i,b=c_i)]  \\
\;\; = \Pr[ j] \cdot \Pr[b = c_j | j] + \sum_{i \neq j} \Pr[i] \cdot \Pr[b = c_i |  i]
\end{align*}
From the analysis of $A_2$, we know that for any single-photon pulse with index $i$, we have $\Pr[b = c_i |  i] \leq y$. Note that again, the state that corresponds to the remaining pulses, including the two-photon one, is independent of $c_i$.
Let $q = \Pr[j]$, then we have
\[
\Pr[x=0|A_4]  = q \cdot \Pr[b = c_j |  j] + (1-q)y
\]
Let us now study the probability that Bob can guess the bit $c_j$ that corresponds to the two-photon pulse. Using the notation from the previous events, the overall state Bob has in case $c_j=0$ is $\sigma_0 \otimes \mathbb{I}_{\neg{j}}$ and the state he has in case $c_j=1$ is $\sigma_1 \otimes \mathbb{I}_{\neg{j}}$. The optimal probability for guessing $c_j$ is given by the optimal POVM on the $K$ pulses that distinguishes the states $\sigma_0 \otimes \mathbb{I}_{\neg{j}}$ and $\sigma_1 \otimes \mathbb{I}_{\neg{j}}$.
This is again given by the Helstrom measurement and has probability $P_{\text{opt}}=\frac{1}{2}+ \frac{1}{2}|| \frac{1}{2}\sigma_0\otimes \mathbb{I}_{\neg{j}} - \frac{1}{2}\sigma_1\otimes \mathbb{I}_{\neg{j}} || = y$.

Let us now describe a specific strategy that Bob can perform in order to guess the value $c_j$ of the two-photon pulse: He performs the POVM  $\{M_{i,b}\}_{i \in [K], b \in \{0,1\}}$ and if the output is $(j,0)$ he outputs 0, if the output is $(j,1)$ he outputs 1, and in all other cases he outputs a uniformly random bit. Let $z$ the success probability of this strategy. Then,
 $z = q \Pr[b = c_j | j] + \frac{1-q}{2}$. This yields the inequality $2z - 1 \le q$. Also, we have $z \le y$, from the optimality of the Helstrom measurement.
This gives:
\begin{align*}\qquad
\lefteqn{\Pr[x=0|A_4]} \\
& \le q  \cdot \Pr[b = c_j |j] + (1-q)y  \\
& = z + (1-q)(y - 1/2) \le z + (2-2z)(y-1/2) \\
& =2z(1-y)+2y-1 \le 2y(1-y)+2y-1 \\
& = -2y^2 + 4y - 1
\end{align*}
\end{description}
By combining the above results, and noticing that Bob's cheating is the same if he wants $x=1$, we find that Bob can bias the coin with probability:
\begin{eqnarray}
p^B_q&\leq&\sum_{i=1}^{4}\Pr[A_i] \cdot \Pr[x=0\lvert A_i]+\Bigg[1-\sum_{i=1}^{4}\Pr[A_i] \Bigg] \cdot 1~~~~~
\label{eq:pqB1}
\end{eqnarray}
As we see from Eqs. (\ref{eq:pqA}) and (\ref{eq:pqB1}), the cheating probabilities depend on the number of rounds $K$, the protocol parameter $y$ and the mean photon number $\mu$. We can make the protocol fair (i.e. $p_q^A = p_q^B$), by changing the parameter $y$.


\subsubsection{Satisfying the security assumptions with the plug and play system.} The previous security analysis holds when the three security assumptions are satisfied. In practice, however, these assumptions, which concern honest Alice and Bob, may not be fulfilled. In the following, we assess the deviations from the security assumptions present in our system and describe the procedures that need to be performed to recover those assumptions almost perfectly and consequently the security of the implemented protocol as well.

In our experimental setup, for every round $i$, honest Alice uses a quantum random number generator to pick the values of the basis $\alpha_i$ and bit $c_i$. She then generates the corresponding state $\ket{\Phi_{\alpha_i,c_i}}$ by applying a suitable phase shift with a phase modulator. On his side, Bob uses a quantum random number generator to pick the basis $\beta_i$, for every $i$, applies the appropriate phase shift with his phase modulator and uses two InGaAs avalanche photodiode single-photon detectors to register the outcome of his measurement. He also uses a quantum random number generator to pick the bit $b$.

We now examine each of the three assumptions we made previously. We will see that the first two assumptions hold almost perfectly since the only deviations come from the possible bias of the quantum random number generator and the variation in the capability of the phase modulator to apply different phase shift values. In order to fulfill the last assumption, however, we need to add to the protocol a symmetrization stage to remedy for an asymmetry in the two detection efficiencies.\\

\paragraph{Assumption 1: Alice's choice of states.} We examine Alice's ability to generate each of the four protocol states with the same probability for each pulse. The only way to assess this in practice is by analyzing Bob's detection events. We calculate the probability that honest Alice had picked each basis over the entire set of Bob's detection events.
The distribution of this basis should ideally be uniform. Based on the experimental data corresponding to the 15 km implementation, we find that the basis choice is close to uniform:
\begin{equation*}
  \begin{split}
    \Pr[\alpha=0] = 0.5048 \\
    \Pr[\alpha=1] = 0.4952
  \end{split}
\end{equation*}
For the 25 km experiment, the corresponding probabilities are 0.5038 and 0.4962, respectively.

Next, we would like to ascertain whether the distribution of Alice's bit for each pulse is also uniform and moreover that it remains uniform even conditioned on Alice's choice of basis. Again we can only assess this by looking at the detection events of Bob. However, it is necessary that we remove the possible effects of Bob's detectors on this distribution, since Bob uses two different detectors for the two measurement outcomes (see Fig. 1 in main text). For this reason, we collected data again from our experiment, where this time, by appropriately adding a phase shift via Bob's phase modulator, we interchanged the role of the two detectors for each of the two bases. The analysis of the obtained data shows that the ratio of detection events corresponding to bits 0 and 1 for each of the bases is almost perfectly inverted. This implies that Alice produces states that correspond to 0 and 1 almost uniformly, even when we condition on her basis choice (with a deviation of $ 0.003$). This is not surprising since the state choice is performed using a quantum random number generator and a single phase modulator that applies one of four possible phase shifts.

By performing extensive tests, we can bound the deviation of Alice's state distribution from the uniform one, using a single $\epsilon_A$:
\COMMENT{
\begin{eqnarray*}
|\Pr[c=k ] - 1/2| \leq 0.0048
\end{eqnarray*}
In addition, using our data, we can calculate that even conditioned on each value of the bit $c$, the probabilities of the two bases are almost uniform. In other words, for all $k,l \in \{0,1\}$
\begin{eqnarray*}
|\Pr[\alpha=k | c=l] - 1/2| \leq 0.0056
\end{eqnarray*}
Hence, we can conclude that in our implementation, we can assume that for all $k,l \in \{0,1\}$,
}
\begin{eqnarray}
\label{eq:epsilonA}
\forall k,l\in \{0,1\}, |\Pr[\alpha=k | c=l] - 1/2|\leq 0.005 \triangleq  \epsilon_A \\
\forall k,l\in \{0,1\}, |\Pr[\alpha=k,c=l] - 1/4| \leq 0.004 \leq \epsilon_A \nonumber\\
\forall k,l\in \{0,1\}, |\Pr[c=l | \alpha=k] - 1/2| \leq 0.003 \leq  \epsilon_A \nonumber
\end{eqnarray}
Since this bound on the deviation is very small, we do not proceed in any correction as part of the experimental procedure, but we incorporate the bound $\epsilon_A$ in our complete security analysis described in a following section. This naturally slightly increases Bob's cheating probability.\\

\paragraph{Assumption 2: Bob's choice of bases and bit $b$.} Next, we examine the probability that Bob has chosen each basis $\beta$ for the first pulse he successfully measured. Note that what is important is the distribution of the basis for the first measured pulse and not over all pulses, since this is the information which is relevant to the coin outcome. The distribution of this basis should ideally be uniform. Note that the same pair of detectors is used for measuring in both bases (see Fig. 1 in main text). From the analysis of the experimental data for the 15 km implementation, we find:
\begin{equation*}
  \begin{split}
    \Pr[\beta=0]=0.5006\\
    \Pr[\beta=1]=0.4994
  \end{split}
\end{equation*}
For the 25 km experiment, the corresponding probabilities are 0.5003 and 0.4997, respectively. The above results demonstrate that the distribution of Bob's bases is indeed very close to uniform. Again, this is not surprising since, as mentioned before, the only devices that are used for the basis and bit choices in our implementation are quantum random number generators and phase modulators, which are expected to be very reliable.

Concerning the distribution of the bit $b$, we note that the quantum random number generator used in our experiment (Quantis) provides very strong guarantees for the uniformity of each output bit and the independence between different output bits.
In fact, we extensively tested the bias of the outputs of Quantis and we can bound the deviation of the probability of each bit from uniform, even conditioned on any number of previous bits, by $\epsilon_Q=0.0006$.

Hence, in our implementation, we can bound, for the first pulse Bob successfully measures, the deviation of his distribution of basis and bit from uniform, as follows:
\begin{equation}
 \forall k,l\in \{0,1\}, |\Pr[\beta=k,b=l]-1/4| \leq 0.00061 \triangleq \epsilon_B
\label{eq:epsilonB}
\end{equation}
Again, since the deviation is very small we do not proceed in any correction in practice, however we incorporate the bound $\epsilon_B$ in our security analysis. This slightly increases Alice's cheating probability.\\

\noindent {\em Assumption 3: Bob's detection.}
In order to calculate the detection efficiency ratio, we focus on the number of detection events that occur when Alice and Bob have used the same bases and agree on the output values (see Suppl. Table \ref{table1}). It is clear that there exists a significant asymmetry in the number of detections observed by each detector, which leads to an important bias in the announced outcomes by Bob. Our previous analysis has practically excluded that this event is due to an imbalance in the states Alice prepares, hence it is predominantly due to an asymmetry in the two detection efficiencies in Bob's system.

After subtracting the events that are due to dark counts from the total number of detection events, taking also into account the slight asymmetry of Bob's choice of basis, we find that the detector efficiency ratio, for both channel lengths, is $\eta_1/\eta_0=0.68\pm0.015$, where $\eta_0$ and $\eta_1$ correspond to detectors $D_0$ and $D_1$, respectively (see Fig. 1 in main text). This difference in detection efficiencies can lead to a sophisticated attack by Alice, where she sends a different state than $\ket{+}$ or $\ket{-}$ (depending on which bit is favored by the asymmetry). This can increase her cheating probability substantially. An efficient solution proposed in Ref. \cite{NJM:natcom12} is the symmetrization of losses, by which Bob effectively makes the two detection efficiencies equal by throwing away some detection events. More specifically, whenever Bob detects an event on detector $D_0$, he discards it with probability 32\%. This was implemented in our experiments. Even after the symmetrization procedure, some deviation on the detection efficiency ratio may still remain, and by testing our detectors can be bounded as follows:
\begin{equation}
\left|\frac{\eta_1}{\eta_0}-1 \right| \leq  0.022 \triangleq \epsilon_B^\prime
\label{eq:epsilonB'}
\end{equation}
The bound $\epsilon_B'$, which again holds with probability negligibly away from 1, is incorporated in the security analysis that follows and increases Alice's cheating probability.

\subsubsection{Security analysis for the plug and play system} We now provide a general security analysis of the basic quantum coin flipping protocol, which incorporates the imperfections quantified by the bounds in Eqs. (\ref{eq:epsilonA}), (\ref{eq:epsilonB}), and (\ref{eq:epsilonB'}), and calculate the cheating probabilities in our implementation.\\\\

\paragraph{Malicious Alice.} Let us assume that Alice tries to bias the coin towards the value $x=0$ (the analysis for $x=1$ is similar) and that Bob successfully measured the first pulse. We assume that the probabilities of Bob's distribution of the basis $\beta$ he chose for the first successfully measured pulse and his bit $b$ deviate at most $\epsilon_B$ from $1/4$. We also assume that Alice has the power to choose among all these distributions the one that maximizes her cheating probability.

As in the uniform case, Alice's optimal strategy consists of finding the state that will maximize the average probability of revealing bit 0 or 1. Even in the presence of the small deviation of Bob's choices, the arguments in Ref. \cite{SR:QIC02} still show that in our protocol, Alice's optimal strategy is to send the state that maximizes the probability of revealing:
\begin{enumerate}
\item $(\alpha=0,c=0)$ when $b=0$ and $(\alpha=1,c=1)$ when $b=1$, or
\item $(\alpha=1,c=0)$ when $b=0$ and $(\alpha=0,c=1)$ when $b=1$.
\end{enumerate}
Note that, due to the deviation $\epsilon_B$, these two optimal strategies may not achieve the same cheating probability, which means that we need to calculate both of them and take the maximum of the two. We remind that in the case of $\epsilon_B=0$, the two optimal strategies correspond to sending the states $\ket{+}$ and $\ket{-}$ and they achieved the same cheating probability.

Let us analyze the first strategy (the analysis of the other one is similar). \COMMENT{ Without loss of generality, Alice will send a pure state:
\[
\ket{\chi}=\cos\phi\ket{\Phi_{0,0}}+\sin\phi\ket{\Phi_{0,1}}
\] } Let $\rho$ the state sent by Alice.
The probability that the protocol outputs $x=0$ is:
\begin{eqnarray*}
\Pr[x=0]&=& \sum_{k,l} \Pr[\beta=k, b=l]\Pr[x=0|\beta=k,b=l]
\end{eqnarray*}
where $\beta$ is Bob's choice of basis and $b$ is Bob's bit, for the successfully measured pulse. According to Alice's strategy, when Bob picks $\beta \neq b$, then Alice reveals a different basis than Bob's, which means he accepts with probability 1. In other words
$$ \Pr[x=0|\beta=0,b=1]=\Pr[x=0|\beta=1,b=0]=1.$$
To upper-bound Alice's cheating probability, we attribute the highest possible probability to these events, more precisely:
\begin{eqnarray*}
\Pr[\beta=0,b=1]=\Pr[\beta=1,b=0] = \frac{1}{4}+\epsilon_B\\
\Pr[\beta=0,b=0]=\Pr[\beta=1,b=1] = \frac{1}{4}-\epsilon_B
\end{eqnarray*}
Then, we need to compute the probability that the protocol outputs 0, when Alice sends the state $\rho$ and Bob picks $\beta=b$ for the successfully measured pulse. Note that, by definition of the first strategy, when $(\beta=0,b=0)$, Alice reveals $(\alpha=0,c=0)$ and when $(\beta=1,b=1)$, she reveals $(\alpha=1,c=1)$. Let us assume that the ratio of the detection efficiencies of Bob's system deviates from 1 by at most $\epsilon_B^\prime$ and Alice knows this distribution. Then, the probabilities are
\begin{align*}
\Pr[x=0|\beta=0,b=0]
&=  \frac{\bra{\Phi_{0,0}} \rho \ket{\Phi_{0,0}}\eta_0 }{\bra{\Phi_{0,0}} \rho \ket{\Phi_{0,0}}\eta_0+ \bra{\Phi_{0,1}} \rho \ket{\Phi_{0,1}} \eta_1}\\
\Pr[x=0|\beta=1,b=1]
&=  \frac{\bra{\Phi_{1,1}} \rho \ket{\Phi_{1,1}} \eta_1 }{  \bra{\Phi_{1,0}} \rho \ket{\Phi_{1,0}} \eta_0+ \bra{\Phi_{1,1}} \rho \ket{\Phi_{1,1}} \eta_1}
\end{align*}\COMMENT{
& \leq & \frac{\sin^2(2\theta+\phi)+\epsilon_B^\prime \sin^2(2\theta+\phi)}{1+\epsilon_B^\prime \sin^2(2\theta+\phi)}
\end{align*}}
We can see that the maximum value of the above expressions, i.e., the maximum of Alice's cheating probability, can be achieved by a pure state that belongs to the Hilbert space defined by the honest states $\ket{\Phi_{\alpha,c}}$. This is the same result as in the uniform case, where the optimum was achieved by the states $\ket{+}$ and $\ket{-}$. Note that the expressions are concave in $\rho$ and the extremal points of the set of density matrices are pure states; in addition, any part of $\rho$ outside the Hilbert space of the honest states leaves the expressions unchanged.

Since $\{ \ket{\Phi_{0,0}} , \ket{\Phi_{0,1}} \}$ is a basis for this space, we have that there exists a state that maximizes Alice's cheating of the form:
\[
\ket{\chi}=\cos\phi\ket{\Phi_{0,0}}+\sin\phi\ket{\Phi_{0,1}}
\]
We can then optimize over all angles $\phi$ to find an upper bound on Alice's cheating probability $p_q^A = \max \{\Pr[x=0],\Pr[x=1]\}$. The analysis for $x=1$ gives the same results. Note that we have upper-bounded the cheating probability by giving Alice knowledge of the efficiency ratio and also the power to attribute in the best way the deviations.

In the case $\epsilon_B=\epsilon_B^\prime=0$, we find that the optimal cheating strategy is for $\phi=\pi/4-\theta$ and we recover Alice's original optimal cheating strategy, which leads to the cheating probability in Eq. (\ref{eq:pqA}).\\\\

\paragraph{Malicious Bob.} Now let us see how cheating Bob can exploit the deviation in Alice's distribution of choices. We assume that Alice's probabilities deviate from uniform by at most $\epsilon_A$ (see Eq. (\ref{eq:epsilonA})).
We analyze the four cheating events for Bob in a very similar way as before. Bob is assumed to want a coin value $x=0$ (the same analysis holds for $x=1$).
\begin{itemize}

\item[$A_1$:] Bob receives only vacuum pulses. Bob picks $b$ equal to the most probable bit according to Alice's distribution. We have $\Pr[x=0|A_1] \leq \frac{1}{2}+\epsilon_A$. Note again that the value of Alice's bit has deviation $\epsilon_A$ from uniform, even conditioned on any of the other bits she has encoded in different pulses.

\item[$A_2$:] Bob receives vacuum pulses, at least one single-photon pulse and no two- or more-photon pulses. Let $\rho_0$ and $\rho_1$ the state of a single-photon pulse corresponding to Alice's bit 0 and 1, in case her distribution is uniform and let $\rho_0^\prime$ and $\rho_1^\prime$ the states when Alice's distribution deviates from uniform by at most $\epsilon_A$. We will show that there exists $y' \in (\frac{1}{2},1)$, such that
\[ \forall m \in \{0,1\},\;\;\; 2(1-y') \mathbb{I} \preceq \rho_{m}^\prime \preceq 2y'\mathbb{I},
\]
Then, we can follow the same analysis as in the uniform case and conclude that
$
\Pr[x=0|A_2]  \leq  y'.
$

To compute $y'$, note that
\begin{eqnarray*}
\rho_0^\prime&=&\Pr[\alpha=0|c=0]\ketbra{\Phi_{0,0}}{\Phi_{0,0}} \\
&+&\Pr[\alpha=1|c=0] \ketbra{\Phi_{1,0}}{\Phi_{1,0}}\\
\rho_1^\prime&=&\Pr[\alpha=0|c=1]\ketbra{\Phi_{0,1}}{\Phi_{0,1}}\\
& +& \Pr[\alpha=1|c=1]\ketbra{\Phi_{1,1}}{\Phi_{1,1}}
\end{eqnarray*}
Since these probabilities deviate at most $\epsilon_A$ from $1/2$, we have for all $m \in \{0,1\}$
\begin{eqnarray*}
(1-2\epsilon_A)\rho_m & \preceq &\rho_m^{\prime} \preceq (1+2\epsilon_A) \rho_m  \Leftrightarrow \\
(1-2\epsilon_A)2(1-y) \mathbb{I} & \preceq& \rho_m^{\prime} \preceq (1+2\epsilon_A) 2y \mathbb{I}  \Leftrightarrow \\
2(1-(y+2\epsilon_A)) \mathbb{I} & \preceq &\rho_m^{\prime} \preceq 2(y+2\epsilon_A)\mathbb{I}
\end{eqnarray*}
Similarly to the analysis of $A_2$ in the uniform case, we have
$\Pr[x=0|A_2]  \leq  y + 2\epsilon_A.$

\item[$A_3$:] Bob receives only vacuum pulses and one two-photon pulse. Let $\sigma_0$ and $\sigma_1$ the mixed states that correspond to the two-photon pulse Bob receives when Alice's bit is 0 or 1 in case Alice's distribution is exactly uniform, and $\sigma_0^\prime $ and $\sigma_1^\prime $ the mixed states that correspond to the two-photon pulse Bob receives when Alice's distribution deviates from uniform by $\epsilon_A$:
\begin{eqnarray*}
\sigma_0^\prime&=&\Pr[\alpha=0|c=0] (\ketbra{\Phi_{0,0}}{\Phi_{0,0}})^{\otimes 2} \\
&+&\Pr[\alpha=1|c=0] (\ketbra{\Phi_{1,0}}{\Phi_{1,0}})^{\otimes 2}\\
\sigma_1^\prime&=&\Pr[\alpha=0|c=1](\ketbra{\Phi_{0,1}}{\Phi_{0,1}})^{\otimes 2}\\
& +& \Pr[\alpha=1|c=1](\ketbra{\Phi_{1,1}}{\Phi_{1,1}})^{\otimes 2}
\end{eqnarray*}
Again, the optimal distinguishing measurement gives:
\begin{eqnarray*}
\lefteqn{\Pr[x=0|A_3]}\\
& = & \frac{1}{2}+\frac{1}{2}\norm{\Pr[c=0]\sigma_0^\prime -\Pr[c=1]\sigma_1^\prime}\\
&\leq&\frac{1}{2}+\frac{1}{2}\norm{\frac{1}{2}\sigma_0-\frac{1}{2}\sigma_1}+\epsilon_A \norm{\sigma_0+\sigma_1}\\
& \leq & y + 2\epsilon_A
\end{eqnarray*}

\item[$A_4$:]
The analysis is similar to the case of the uniform distribution. Bob receives vacuum pulses, one two-photon pulse and at least one single-photon pulse. This events occurs with probability $\Pr[A_4]=Kp_2[(p_0+p_1)^{K-1}-p_0^{K-1}]$. We will assume here as before that Bob receives no vacuum pulses, which can only increase his cheating.

Let $\{M_{i,b}\}_{i \in [K], b \in \{0,1\}}$ be the POVM that Bob applies on all $K$ pulses to determine the index $i$ Alice and Bob will use as well as his guess $b$ for Alice's bit $c_i$. Let $j$ the index that corresponds to the two-photon pulse. We have
\begin{align*}
&~~ \Pr[x=0|A_4] \\
&~~~~~~~ = \sum_{i \in [K]} \Pr[\mbox{Bob outputs} (i,b=c_i)]  \\
&~~~~~~~ = \Pr[ j]\Pr[b = c_j | j] + \sum_{i \neq j} \Pr[i] \cdot \Pr[b = c_i |  i]
\end{align*}
From the analysis of $A_2$, we know that for any single-photon pulse with index $i$, we have $\Pr[b = c_i |  i] \leq y+2\epsilon_A$.
Let $q = \Pr[j]$, then we have
\[
\hspace{0.4in}\Pr[x=0|A_4]  = q \cdot \Pr[b = c_j |  j] + (1-q)(y+2\epsilon_A)
\]
Let us now study the probability that Bob can guess the bit $c_j$ that corresponds to the two-photon pulse. Using the notation from the previous events, the overall state Bob has in case $c_j=0$ is $\sigma_0^\prime \otimes \gamma_{\neg{j}}$ and the state he has in case $c_j=1$ is $\sigma_1^\prime \otimes \gamma_{\neg{j}}$ for some state $\gamma_{\neg{j}}$.
The optimal probability for guessing $c_j$ is given by the optimal POVM on the $K$ pulses that distinguishes the states $\sigma_0^\prime \otimes \gamma_{\neg{j}}$ and $\sigma_1^\prime \otimes \gamma_{\neg{j}}$.
This is again given by the Helstrom measurement and gives probability $P_{\text{opt}} = \frac{1}{2}+ \frac{1}{2}|| \Pr[c=0]\sigma_0^\prime \otimes \gamma_{\neg{j}} - \Pr[c=1] \sigma_1^\prime \otimes \gamma_{\neg{j}} || \leq y + 2\epsilon_A$.

Let us now describe a specific strategy that Bob can perform in order to guess the value $c_j$ of the two-photon pulse: He performs the POVM  $\{M_{i,b}\}_{i \in [K], b \in \{0,1\}}$ and if the output is $(j,0)$ he output 0, if the output is $(j,1)$ he outputs 1, and in all other cases he outputs the most probable value. Let $z$ the success probability of this strategy. Then,
 $z = q \Pr[b = c_j | j] + (1-q)(\frac{1}{2}+\epsilon_A)$. This yields the inequality $\frac{2z - 1-2\epsilon_A}{1-2\epsilon_A} \le q$. Also, we have $z \le y+2\epsilon_A$, from the optimality of the Helstrom measurement.
This gives us:
\begin{eqnarray*}\qquad
\lefteqn{\Pr[x=0|A_4]}\\
 & = q  \cdot \Pr[b = c_j |j] + (1-q)(y+2\epsilon_A)  \\
& = z + (1-q)(y - 1/2+\epsilon_A) \\
& \le z + (1-\frac{2z - 1-2\epsilon_A}{1-2\epsilon_A} )(y-1/2+\epsilon_A) \\
& = \frac{2}{1-2\epsilon_A} \left(z(1-y-2\epsilon_A)+y-\frac{1}{2}+\epsilon_A\right)
\end{eqnarray*}
Since the coefficient of $z$ is positive for the values of $y$ and $\epsilon_A$ that we consider, we can upper bound this probability by using $z\leq y+2\epsilon_A$ and get:
\begin{align*}
~~~~~~~\Pr[x=0|A_4] \leq \frac{(-2y^2+4y-1)+\epsilon_A(6-8y-8\epsilon_A)}{1-2\epsilon_A}
\end{align*}
Note that for $\epsilon_A=0$, we recover the initial bound.
\end{itemize}
By combining the above results, and noticing that Bob's cheating is the same if he wants $x=1$, we find that Bob can bias the coin with probability:
\begin{eqnarray}
p^B_q&\leq&\sum_{i=1}^{4}\Pr[A_i] \cdot \Pr[x=0\lvert A_i]+\Bigg[1-\sum_{i=1}^{4}\Pr[A_i] \Bigg] \cdot 1~~~~~
\end{eqnarray}

By inserting the specific values for $\epsilon_A, \epsilon_B,\epsilon_B^\prime$ from our implementation, given in Eqs. (\ref{eq:epsilonA}), (\ref{eq:epsilonB}), and (\ref{eq:epsilonB'}), we calculate the cheating probability of the protocol. For a channel length of 15 km there is a clear quantum gain. For 25 km, we see that the quantum cheating probability, taking into account the worst possible case for all the imperfections, is slightly bigger than the classical bound, and hence we cannot provably show any gain in this case (see Fig. 2 in main text).

An important part of our analysis is the symmetrization procedure that results in throwing away a lot of detection events, thus requiring more rounds $K$ in order to achieve a specific honest abort probability. This  increases the cheating probability since a malicious Bob benefits from the increased number of rounds.


\subsection{Supplementary Note 2 - Combined quantum coin flipping protocols}

In this section, we show how to combine the basic quantum coin flipping protocol with protocols that achieve almost perfect security against adversaries that possess limited resources. We consider, in particular, computationally bounded adversaries and adversaries with noisy quantum storage.

\subsubsection{Computationally-bounded quantum coin flipping} The computationally bounded protocol shown in Supplementary Protocol \ref{compProtocol}, uses an injective one-way function $f$, upon which Alice and Bob have previously agreed \cite{goldreich01}. In the \emph{commit} stage of the protocol, Alice and Bob choose random strings, $x_A$ and $x_B$, respectively, and commit to the bits $h(x_A)$ and $h (x_B)$ by exchanging $f(x_A)$ and $f(x_B)$, where $h$ is a hardcore predicate of $f$. Hardcore predicates make it impossible to guess the value $h(x)$ from $f(x)$ with probability greater than one half. An  example of a hardcore predicate function is the parity of the bits in a string, since it can be proven \cite{goldreich01} that given the parity and the image of the string, it is not feasible to guess the string itself. Moreover, since $f$ is an injective one-way function, by sending the values $f(x)$, neither of the two parties can lie about the value of their chosen string and thus change the value $h(x)$. Hence, at the end of this stage Alice and Bob have almost perfectly committed to $h(x_A)$ and $h(x_B)$. In the \emph{encrypt} stage, for $i=1,...,K$, Alice randomly selects $\alpha_i$ and $c_i$ and sends the $K$ quantum states $\ket{\Phi_{\alpha_i,c_i\oplus h(x_A)}}$ to Bob, prepared in the same way as in the basic protocol. Bob performs a measurement in the randomly selected bases $\{\ket{\Phi_{\beta_i,0}},\ket{\Phi_{\beta_i,1}}\}$, and replies with the position $j$ of the first successfully measured pulse and a random bit $b$ encrypted as $b\oplus h(x_B)$. Finally, in the \emph{reveal} stage, Alice and Bob reveal their strings and check that they are consistent with the function outputs exchanged during the \emph{commit} phase. They also exchange their chosen bit and Bob aborts only if $\alpha_j=\beta_j$ and his measurement outcome does not agree with $c_j$. If he does not abort, the value of the coin is $c_j\oplus b$. Note that the \emph{encrypt} stage and the first step of the \emph{reveal} stage correspond to our basic quantum coin flipping protocol, slightly modified to fit the underlying computationally bounded model.

For the security analysis, if Alice is computationally bounded, then she cannot guess the value $h(x_B)$ with probability greater than one half, which means that Bob's bit $b$ is perfectly hidden when Bob sends $b \oplus h(x_B)$. Therefore, the protocol remains almost perfectly secure against Alice. If Bob is computationally bounded, then the bits $c_j$ are perfectly hidden as $c_j \oplus h(x_A)$, hence the protocol remains almost perfectly secure against Bob. If, on the other hand, the parties are unbounded, they can perfectly compute the hardcore predicates and the security of the protocol becomes exactly the same as the security of our basic coin flipping protocol.

\begin{supprotocol}
\begin{adjustbox}{minipage=1\linewidth,fbox,center}
\begin{tabular}{ccc}
\bfseries \underline{Alice}  &  &\bfseries \underline{Bob}\\
choose $x_A$                          &            $\xrightarrow{\makebox[1.8cm]{$\scriptstyle f(x_A)$}}$  &         choose $x_B$                  \\\vspace{0.15in}
                                           &            $\xleftarrow{\makebox[1.8cm]{$\scriptstyle f(x_B)$}}$    &              \\
~~choose $\{\alpha_i,c_i\}_1^K$~~~~~  &    $\xrightarrow{\makebox[1.8cm]{$\scriptstyle \ket{\Phi_{\alpha_i,c_i\oplus h(x_A)}}$}}$     &  measure in $\{\beta_i\}_1^K$\\
                          & $\xleftarrow{\makebox[1.8cm]{$\scriptstyle j,~b\oplus h(x_B)$}}$  &     $j$: first measured pulse, \\
                           &                                                                         &     $b\in_R\{0,1\}$\\
                         & $\xrightarrow{ \makebox[1.8cm]{$\scriptstyle x_A,~c_j,~\alpha_j$}}$ &  \\
                         & $\xleftarrow{\makebox[1.8cm]{$\scriptstyle  x_B,~b  $}}$ &  \\
&\bfseries Coin: $c_j\oplus b$&
\end{tabular}
\end{adjustbox}
\caption{\textbf{Computationally-bounded coin flipping.}}\label{compProtocol}
\end{supprotocol}

\subsubsection{Noisy storage quantum coin flipping} In the noisy storage protocol \cite{NJM:natcom12}, shown in Supplementary Protocol \ref{noisyProtocol}, the parties first agree on an error-correcting code. This is followed by a \emph{prepare} stage, where Alice sends to Bob $2n$ quantum states, which are the states used in the basic protocol, with $y = 1/2$. Bob measures the states using randomly chosen bases $\{\hat{b}_i\}_1^{2n}$. At the end of this procedure, Alice has a string containing the bits used to construct the states, namely $X^{2n} = X_1^n X_2^n$, and Bob has a string containing his measurement results, namely $\tilde{X}^{2n} = \tilde{X}_1^n \tilde{X}_2^n$. If the choices of the states and the measurement bases are uniformly random, then the strings agree on approximately half of the positions.

The parties then perform the main coin flipping protocol. In the $\emph{commit}$ stage, Alice and Bob commit to bits $D_A = \mbox{Ext}(X_1^n,r)$ and $D_B = \mbox{Ext}(\tilde{X}_2^n,\tilde{r})$, respectively, where $\mbox{Ext}:\{0,1\}^n \otimes R \rightarrow \{0,1\}$ is a family of 2-universal hash functions, and ($r,\tilde{r}$) are strings chosen by Alice and Bob in order to randomly pick a hash function from this family. To this end, they first calculate the syndromes $w=Syn({X}_1^n)$ and $\tilde{w}=Syn(\tilde{X}_2^n)$ based on the chosen error-correcting code, and commit to the extractor function values by exchanging the syndromes and half of the bases' values they used in the measurements. In the \emph{encrypt} stage, Alice encrypts her bit choices $c_j$ by sending $K$ states $\ket{\Phi_{\alpha_j,c_j\oplus D_A}}$, prepared as in the basic protocol. Bob chooses randomly $\beta_j$ and measures in $\{\ket{\Phi_{\beta_j,0}},\ket{\Phi_{\beta_j,1}}\}$. He then encrypts a bit $b$ by sending $b\oplus D_B$ to Alice, together with the index $m$ of the first successfully measured pulse. Finally, in the \emph{reveal} stage, Alice and Bob reveal their string and bit values and check that for the positions with the same bases, $X_2^n$ coincides with $\tilde{X}_2^n$ and $X_1^n$ coincides with $\tilde{X}_1^n$, respectively. They also check that the  syndromes and extractor outputs correspond to the received strings.

\begin{supprotocol}[h!]
\begin{adjustbox}{minipage=1\linewidth,fbox,center}
\begin{tabular}{ccc}
\bfseries \underline{Alice}  &  &\bfseries \underline{Bob}\\
choose  $\{b_i,x_i\}_1^{2n}$  & $\xrightarrow{\makebox[1.8cm]{$ \scriptstyle \ket{\Phi_{b_i,x_i}}$}}$     &  measure in $\{\hat{b}_i\}_1^{2n}$\\
String: $X^{2n}=X_1^nX_2^n$& &String:   $\tilde{X}^{2n}=\tilde{X}^n_1\tilde{X}^n_2$\\
$|X^n_1|=|X^n_1|=n$,    &   &   $|\tilde{X}^n_1|=|\tilde{X}^n_1|=n$\\
\multicolumn{3}{c}{\boxed{~~~~~~~~~~~~~~~~~~~~~~~~~~~~\text{Wait time $\Delta t$}~~~~~~~~~~~~~~~~~~~~~~~~~~~~~~}}\\

$w=Syn(X^n_1),$            &                  &                        $\tilde{w}=Syn(\tilde{X}^n_2),$\\
string $r$                          &                     &                        string $\tilde{r}$\\
$D_A=Ext(X^n_1,r)$      & $\xrightarrow{\makebox[1.8cm]{$\scriptstyle \{b_i\}_1^{n/2},~w,~r$}}$                                    &   $D_B=Ext(\tilde{X}^n_2,\tilde{r})$\\
                                         & $\xleftarrow{\makebox[1.8cm]{$\scriptstyle \{\tilde{b}_i\}_{n/2+1}^n,~\tilde{w},~\tilde{r}$}}$         &  \\\\

choose   $\{\alpha_j,c_j\}_1^K$     & $\xrightarrow{\makebox[1.8cm]{$\scriptstyle \ket{\Phi_{\alpha_j,c_j\oplus D_A}}$}}$              &  measure in $\{\beta_j\}_1^K$\\
                                     & $\xleftarrow{\makebox[1.8cm]{$\scriptstyle D_B\oplus b,~m$}}$   &$m$: first measured\\
                                     & & pulse \\
                         & $\xrightarrow{\makebox[1.8cm]{$\scriptstyle X^n_1,c_m$}}$ &  \\
    check:                     & $\xleftarrow{\makebox[1.8cm]{$\scriptstyle \tilde{X}^n_2,~b$}}$ & check: \\
$Syn(\tilde{X}^n_2)=\tilde{w}$  &   &$Syn(X^n_1)=w$\\
$Ext(\tilde{X}^n_2,\tilde{r})=D_B$    && $Ext(X^n_1,r)=D_A$\\
&\bfseries Coin: $c_m\oplus b$&
\end{tabular}
\end{adjustbox}
\caption{\textbf{Noisy storage coin flipping.}}\label{noisyProtocol}
\end{supprotocol}

If the measurement outcome for the first measured pulse agrees with the revealed bit for the same choice of bases or if the bases are different, they agree on the coin, otherwise they abort. Again, the \emph{encrypt} stage and the first step of the \emph{reveal} stage correspond to the basic quantum coin flipping protocol.

The noisy storage limitation together with the waiting time $\Delta t$ that is imposed on the parties, forces them to measure any quantum state they might have wanted to keep unmeasured in order to improve their cheating strategy. Bob is forced to measure the states sent by Alice and Alice is forced to measure whatever entangled share she may have kept when sending the states to Bob.

Concerning the security analysis, if Alice has noisy storage, then she cannot guess the value $D_B$ with probability greater than one half, hence Bob's bit $b$ is perfectly hidden from her when Bob sends $b \oplus D_B$. Therefore, the protocol remains almost perfectly secure against Alice. If Bob has noisy storage, then again the bits $c_j$ are perfectly hidden as $c_j \oplus D_A$ and the protocol remains almost perfectly secure against Bob. If, on the other hand, the parties have perfect memory, they can perfectly compute the values $D_A$ and $D_B$ and the security of the protocol reduces exactly to the security of our basic quantum coin flipping protocol.


\end{document}